\documentclass{article}
\usepackage{graphicx} 
\usepackage{amsmath}
\usepackage{mathrsfs}
\usepackage{authblk}
\usepackage{booktabs}
\usepackage{subfigure}
\usepackage{caption}
\usepackage{hyperref}
\usepackage{float}
\usepackage[normalem]{ulem}
\usepackage{multirow}
\usepackage{arydshln}
\usepackage{geometry}
\usepackage[title]{appendix}

\title{Seamless and multi-resolution energy forecasting}

\date{}

\author[1,2]{Chenxi Wang}
\author[2,3,4,*]{Pierre Pinson}
\author[1, *]{Yi Wang}
\affil[1]{Department of Electrical and Electronic Engineering, The University of Hong Kong}
\affil[2]{Dyson School of Design Engineering, Imperial College London}
\affil[3]{Department of Technology, Management and Economics, Technical University of Denmark}
\affil[4]{Halfspace, Denmark}
\affil[*]{Corresponding authors}

\begin{document}

\maketitle
\begin{abstract}

Energy forecasting is pivotal in energy systems, by providing fundamentals for operation, with different horizons and resolutions. Though energy forecasting has been widely studied for capturing temporal information, very few works concentrate on the frequency information provided by forecasts. They are consequently often limited to single-resolution applications (e.g., hourly). Here, we propose a unified energy forecasting framework based on Laplace transform in the multi-resolution context. The forecasts can be seamlessly produced at different desired resolutions without re-training or post-processing. Case studies on both energy demand and supply data show that the forecasts from our proposed method can provide accurate information in both time and frequency domains. Across the resolutions, the forecasts also demonstrate high consistency. More importantly, we explore the operational effects of our produced forecasts in the day-ahead and intra-day energy scheduling. The relationship between (i) errors in both time and frequency domains and (ii) operational value of the forecasts is analysed. Significant operational benefits are obtained.
\end{abstract}


\section{Introduction}
Energy forecasting is crucial in all segments of the energy industry. The primary goal of energy forecasting is to provide accurate information about both demand \cite{staffell2023global} and supply \cite{alova2021machine, schiermeier2016germany} in the future, such as electricity load and wind power generation. These forecasts support decision-makers to act optimally \cite{hong2020energy}. In general, energy forecasting is a time-series prediction problem in which predictions of diverse temporal horizons and resolutions are required for different applications. Here, forecast horizon refers to how far into the future the forecasts are for, while resolution is defined as the time interval between two successive forecast points. For instance, day-ahead energy demand and renewable energy forecasts with hourly or 15-minute resolutions are used by system operators to schedule energy generation in advance \cite{hobbs1999analysis}. Hours-ahead forecasts at minute resolution are needed for (near) real-time energy balancing \cite{tawn2022review}. And, forecasts at second resolution levels (nowcasts) can be used for immediate control actions and grid stability analysis \cite{pombo2021multi}. Therefore, energy forecasting at different horizons and resolutions is fundamental to all aspects of the operation and a management of the entire energy system.
At one single resolution level, the energy forecasting problem has been widely studied in the context of both energy demand \cite{suganthi2012energy} and supply \cite{wang2019review}. Various approaches for either single-step or multi-step forecasting situations can also be easily found in existing literature reviews \cite{petropoulos2022forecasting} and books \cite{kariniotakis2017renewable}. Despite the prosperity of single-resolution energy forecasting, most of the approaches merely focus on developing a mapping function between relevant temporal features and forecasting targets. However, such methods may be unsuitable for multi-resolution energy forecasting. This is since the frequency information, indicating how intense fluctuations are at various frequencies, may vary depending upon the desired forecasting resolution. Generally, the higher the resolution is, the more detailed frequency information within the energy data should be exhibited \cite{li2021characterizing}. The learned forecasting model for one single resolution can overfit/underfit at other resolutions. Therefore, it may fail to capture the precise frequency information at multiple resolutions. 



Even though multiple forecasting models are trained individually for different resolutions \cite{amara2023daily, liu2020wind}, there still exists challenges in terms of consistency within the temporal hierarchy of the forecasts \cite{athanasopoulos2017forecasting}. For instance, hourly wind power measurements can be readily deduced by averaging minute-resolution observations over 60-minute periods. From both theoretical and practical points of view, if we forecast at both these resolutions, the forecasts generated should also respect this temporal hierarchy. Multiple models, however, by being naturally unaware of this temporal hierarchy, produce forecasts that are likely to be inconsistent. Therefore, for multi-resolution forecasting, such approaches are no longer suitable. In addition, by changing forecast resolution depending upon user needs, the forecasts need to seamlessly adapt, while also respecting underlying temporal hierarchies.

Some previous works aimed at tackling the above issue. For instance they focus on the post-hoc coordination of multi-resolution forecasts \cite{yang2017reconciling, nystrup2020temporal}, the improved training process of multiple models \cite{nejati2023new, liu2020wind2}, and continuous-time neural networks for time series \cite{chen2018neural,hasani2022closed,kidger2020neural}. 
These approaches either try to coordinate the multiple models after/during the training, or purely focus on the dynamics of the time series. They still neglect to reveal the relationship between forecasting resolutions and frequency information. Therefore, from the perspective of methodology, a unified energy forecasting model is needed to seamlessly output the hierarchical forecasts at different resolutions with the corresponding frequency information. 


More importantly, energy forecasts with precise frequency information are supposed to have prominent effects on the following decision-making process \cite{anvari2022data}, especially in situations of high resolution. The connection between temporal forecasting errors and resulting operation results is gradually receiving attention \cite{zhao2021cost,zhang2022cost,li2016toward}, but seldom do the existing works explore the effects of errors in the frequency domain. Consequently, from the perspective of empirical analysis, the final operational effects of the forecasts in terms of both time and frequency deserve to be unveiled.

To this end, we propose an innovative and unified energy forecasting framework, inspired by the recently proposed Neural Laplace framework \cite{holt2022neural}, but with the aim to perform seamless and multi-resolution energy forecasting. We refer to it as Hierarchical Neural Laplace (HNL). Given the desired resolutions, the corresponding forecasts can be seamlessly generated without re-training or post-processing. At the desired resolutions, these forecasts can provide high-quality information in both time and frequency domains accordingly. Across the resolutions, the consistent forecasts comply with temporal hierarchies. We also explore the operational effects of our forecasts in representative decision-making processes for energy scheduling. Our analysis of the relationship between temporal errors, frequency errors, and operational costs reveals significant operational benefits.

\section{Main}
\subsection{Flexible forecasting framework}
\begin{figure}[htb]
    \centering
    \includegraphics[width=\textwidth]{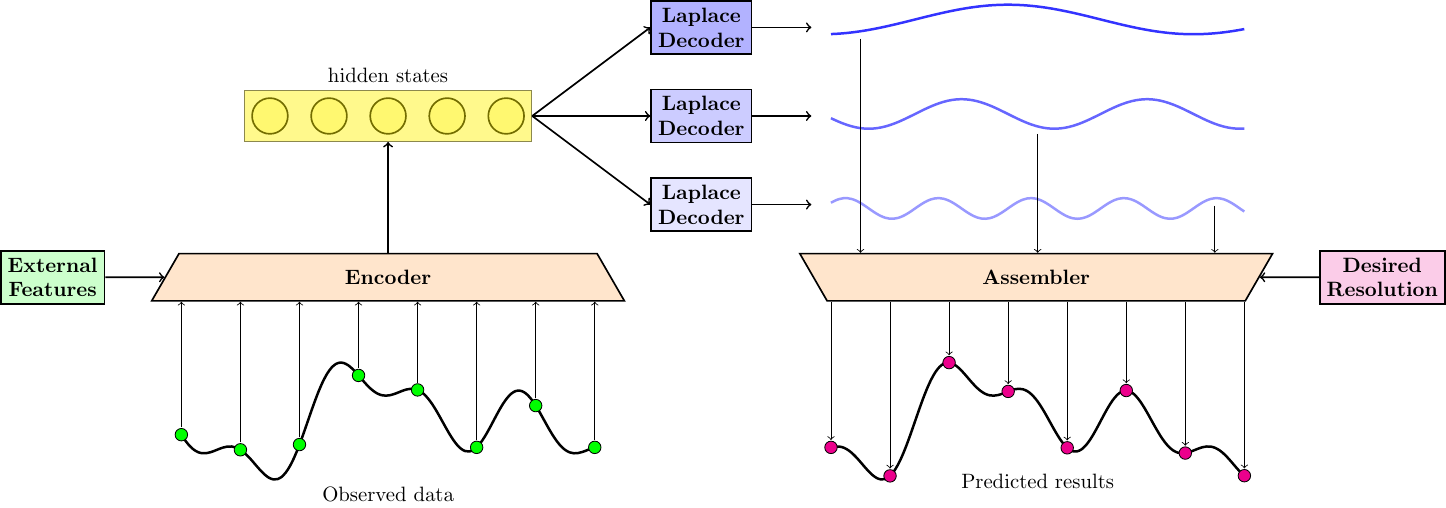}
    \caption{The overview of the proposed Hierarchical Neural Laplace framework}
    \label{fig:frame}
\end{figure}

We briefly introduce the proposed flexible forecasting framework, HNL, in this section. As shown in Fig.~\ref{fig:frame}, it takes the observed energy data at the highest resolution and external features like numerical weather predictions (NWPs) as inputs. Then, temporal components containing different frequency information are produced. Final energy forecasts can be formed seamlessly at the desired resolutions. 

This framework has three core segments: an encoder, multiple Laplace decoders, and an assembler. Firstly, the encoder is responsible for transforming observed data and external features into the high-dimensional representation, i.e., the hidden states. The encoder is model-agnostic, and arbitrary neural networks can be used as encoder, for example, Recurrent Neural Networks (RNN), Convolutional Neural Networks (CNN), etc.

Then, the encoded hidden states will go through multiple Laplace decoders and produce different temporal components. The Laplace transform converts a temporal function into a Laplace function, which illustrates the frequency information in the complex frequency domain. We use different neural networks to fit the Laplace functions on consecutive frequency bands and denote them as Laplace decoders. Inspired by the previous work \cite{holt2022neural}, the Fourier-based discrete Inverse Laplace Transform (ILT) is then utilized to decode these fitted Laplace functions back to temporal functions. They produce temporal components containing different frequency information, as for the blue curves in Fig.~\ref{fig:frame}. Through the analysis of Fourier-based discrete ILT, we can explicitly set the frequency bands for each decoder to learn. Thus, one can learn appropriate representations for chosen ranges of frequencies and yield controllable frequency learning.


Finally, given the desired forecasting resolutions, the assembler will fetch groups of temporal components to form the forecasts with precise frequency information. The Shannon sampling theorem, a bridge between data resolution and frequency information, will guide the assembler to decide which groups of components need to be fetched.

In this way, HNL only needs to be trained once on energy data at the highest resolution, and flexibly generates the energy forecasts at any resolution that we are interested in. Details of the framework can be referred to in Section \ref{chap:method}.

\subsection{Accurate forecasts at different resolutions}
We focus on multi-resolution forecasting case studies on both electricity load data \cite{meinrenken2020mfred} and wind power data \cite{draxl2015wind}, along with NWP data from ECMWF \cite{noauthor_mars_2018}. For each dataset, we produce forecasts at three commonly used resolutions, namely 5-minute, 15-minute, and 60-minute, for the following 24 hours. Prevailing ML-based energy forecasting models, i.e., Multi-Layer Perceptron (MLP) and Long Short-Term Memory (LSTM), are included as benchmarks. We also include the original Neural Laplace (NL) \cite{holt2022neural} as the benchmark. It only has one Laplace decoder. One naive model, Persistence, is included as the baseline model. The latest observed data is used as forecast, for all lead times. Besides, two widely used coordination strategies for multi-resolution forecasts \cite{athanasopoulos2017forecasting}, i.e., bottom-up-based strategy (BU) and optimized-based strategy (OPT), are also applied to each of the benchmark models. Therefore, for each resolution, we have three types of benchmark forecasts:
\begin{itemize}
    \item Raw forecasts from: MLP, LSTM, NL, and Persistence,
    \item BU-based coordinated forecasts from: MLP-BU, LSTM-BU, and NL-BU,
    \item OPT-based coordinated forecasts from: MLP-OPT, LSTM-OPT, and NL-OPT
\end{itemize}
Details of these benchmarks and baseline models can be found in Section 4.

\begin{figure}[H]
\centering  
\subfigure[Comparison on time domain (electricity load)]{   
\begin{minipage}{0.476\textwidth}
\centering    
\includegraphics[width=\textwidth]{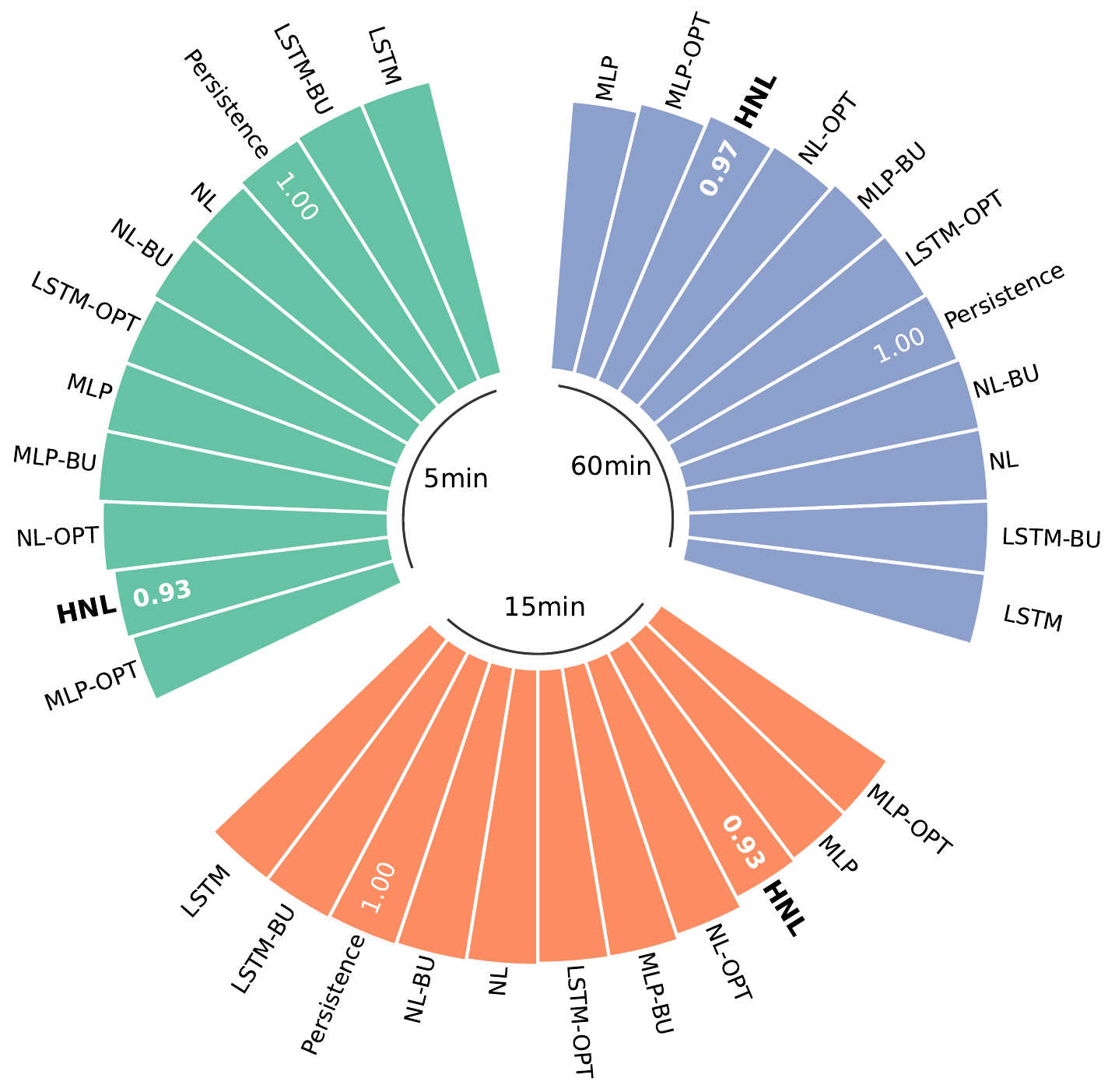}  
\end{minipage}
}
\subfigure[Comparison on frequency domain (electricity load)]{ 
\begin{minipage}{0.476\textwidth}
\centering    
\includegraphics[width=\textwidth]{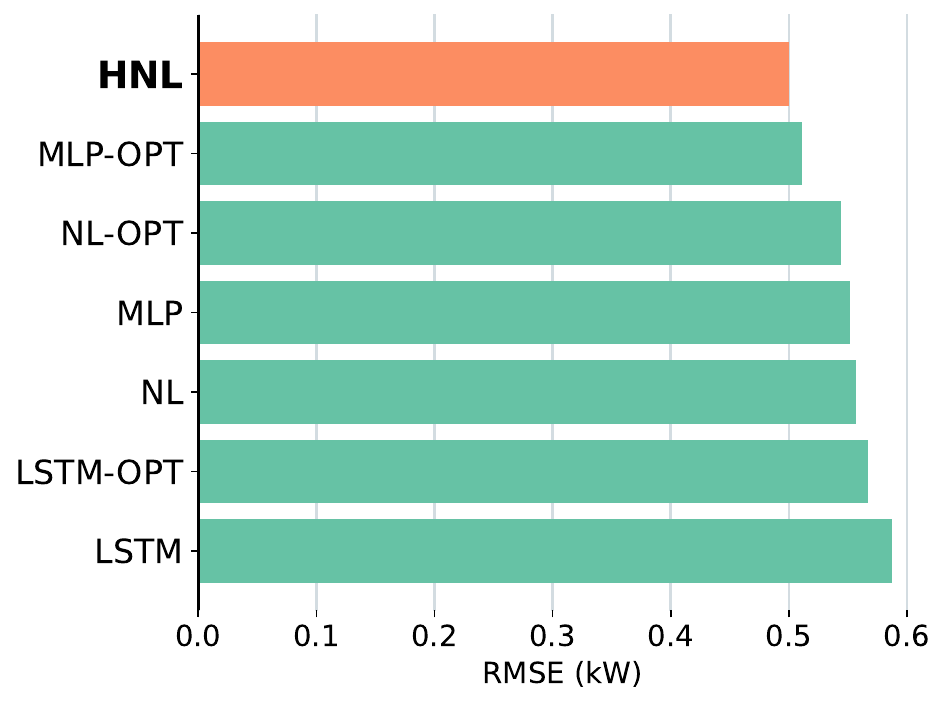}
\end{minipage}
}\\
\subfigure[Comparison in the time domain (wind power)]{   
\begin{minipage}{0.476\textwidth}
\centering    
\includegraphics[width=\textwidth]{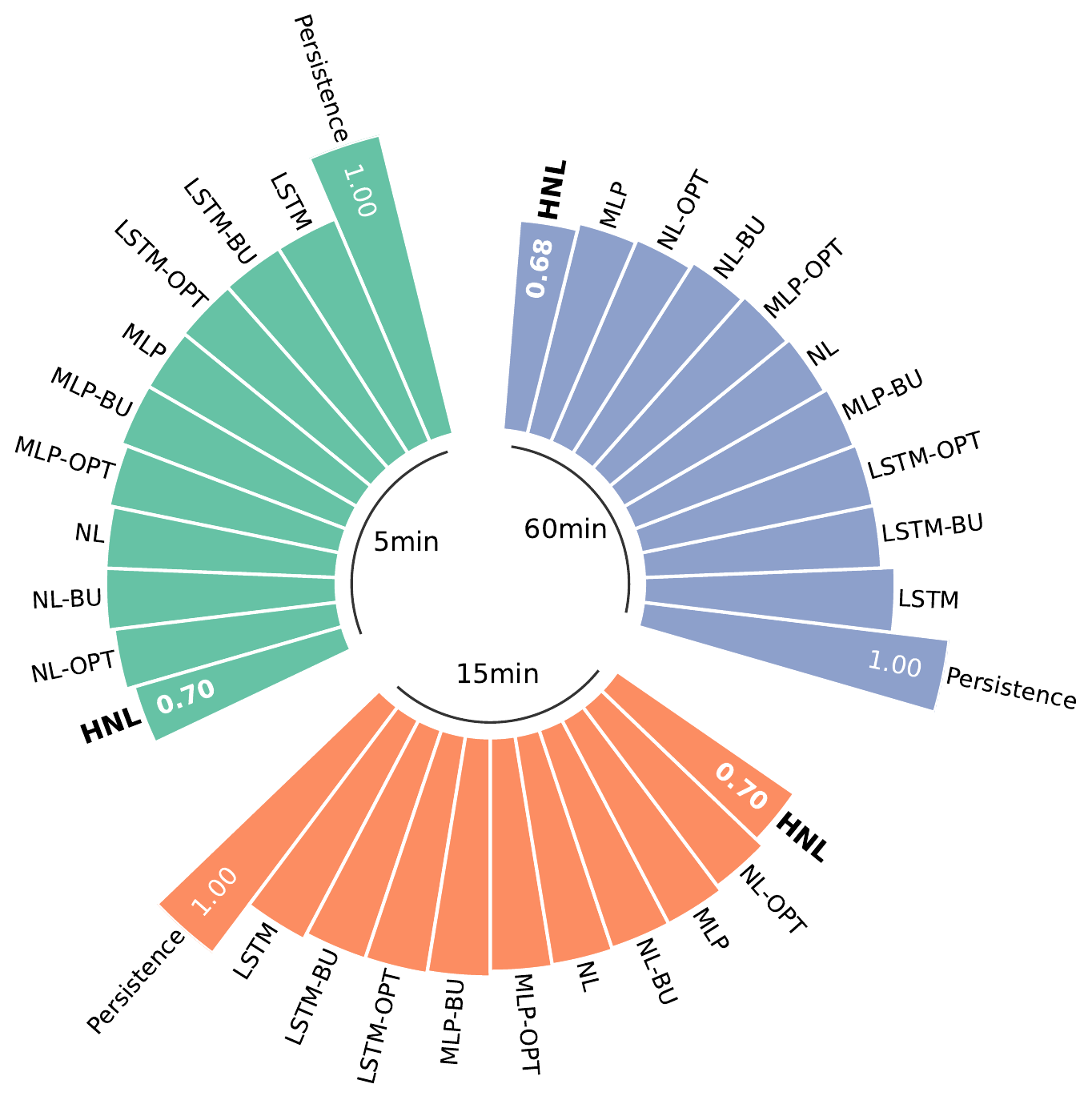}  
\end{minipage}
}
\subfigure[Comparison in the frequency domain (wind power)]{ 
\begin{minipage}{0.476\textwidth}
\centering    
\includegraphics[width=\textwidth]{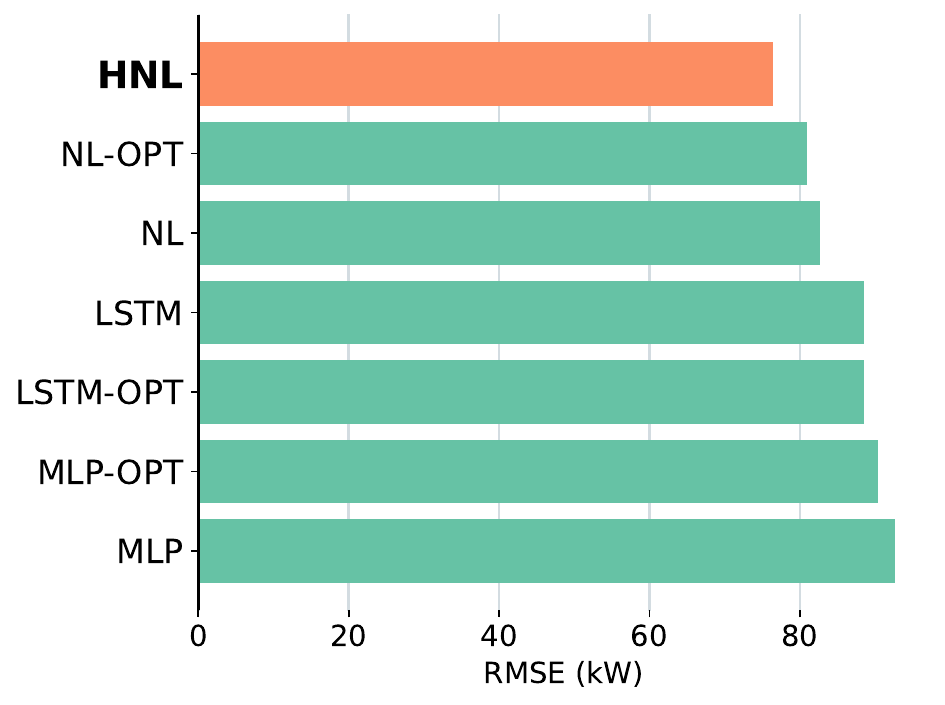}
\end{minipage}
}
\caption{Comparison of forecasting errors (RMSE) in both time and frequency domain in terms of electricity load and wind power. (a) and (c) show the comparison in the time domain, while (b) and (d) show the comparison in the frequency domain at the highest resolution. For (a) and (c), RMSEs are scaled based on the performance of the Persistence model. For (b) and (d), BU-based forecasts are omitted because they are the same as the raw forecasts at the highest resolution.}    
\label{fig:accuracy}    
\end{figure}

To comprehensively evaluate forecast performance, we run each forecast model 20 times with different random seeds. Then, the Root Mean Squared Error (RMSE) at each and every resolution is calculated for each seed, and the comparison of average RMSEs over all seeds is depicted in Fig.~\ref{fig:accuracy}(a)(c).

For electricity load forecasting, the proposed HNL framework outperforms the baseline model by 7.28\%, 7.08\%, and 3.4\%, respectively, at 5-minute, 15-minute, and 60-minute. Coordinating strategies, especially OPT-based strategy, generally lower the forecasting errors of benchmarks at all resolutions. 
Compared to the benchmarks, HNL still shows superiority over most of them and achieves competitive performance with the state-of-the-art coordinated model (MLP-OPT). For wind power forecasting, the performance differences are more remarkable. Outperforming the baseline Persistence by at least 29\%, the proposed HNL obtains the lowest forecasting error at all resolutions. It still has at least a 3\% accuracy improvement over the best-performing benchmarks at each resolution. Thus, on both load and wind power forecasting, the competitive performance of HNL is obvious in terms of forecast quality in the time domain at all resolutions.

In addition to the performance in the time domain, we also show that our proposed method can capture more precise information in the frequency domain. We calculate the RMSE in the frequency domain at the highest resolution (detailed in Section~\ref{chap:method}). In Fig.~\ref{fig:accuracy}(b)(d), HNL attains the best performances in the frequency domain for both datasets, owing to the hierarchical frequency learning. With the limited capability of frequency learning, there is at least 5\% performance lag for NL-based methods. Despite the close performance of MLP-OPT with HNL in the load forecasting, MLP-based models fail to accurately portray the actual frequency information of wind power. Similarly, mainly focusing on the temporal dependence, LSTM-based models cause around 15\% more error in the frequency domain than our proposed HNL. Therefore, besides the competitive forecasting performance in the time domain, our proposed method can capture the information in the frequency domain more accurately than the prevailing forecasting models. 

\subsection{Consistent forecasts across resolutions}
Beyond the accuracy at each resolution, we also evaluate the consistency of the forecasts across these resolutions. The Mean Consistency Error (MCE) is computed as the distance between two forecasts at different resolutions, for example, 5-minute v.s.\ 15-minute. The Total Consistency Error (TCE) is summed by the MCE of each resolution pair. The detailed definitions of MCE and TCE can be found in Section \ref{chap:method}. 

\begin{figure}[htb]
    \centering
    \includegraphics[width=\textwidth]{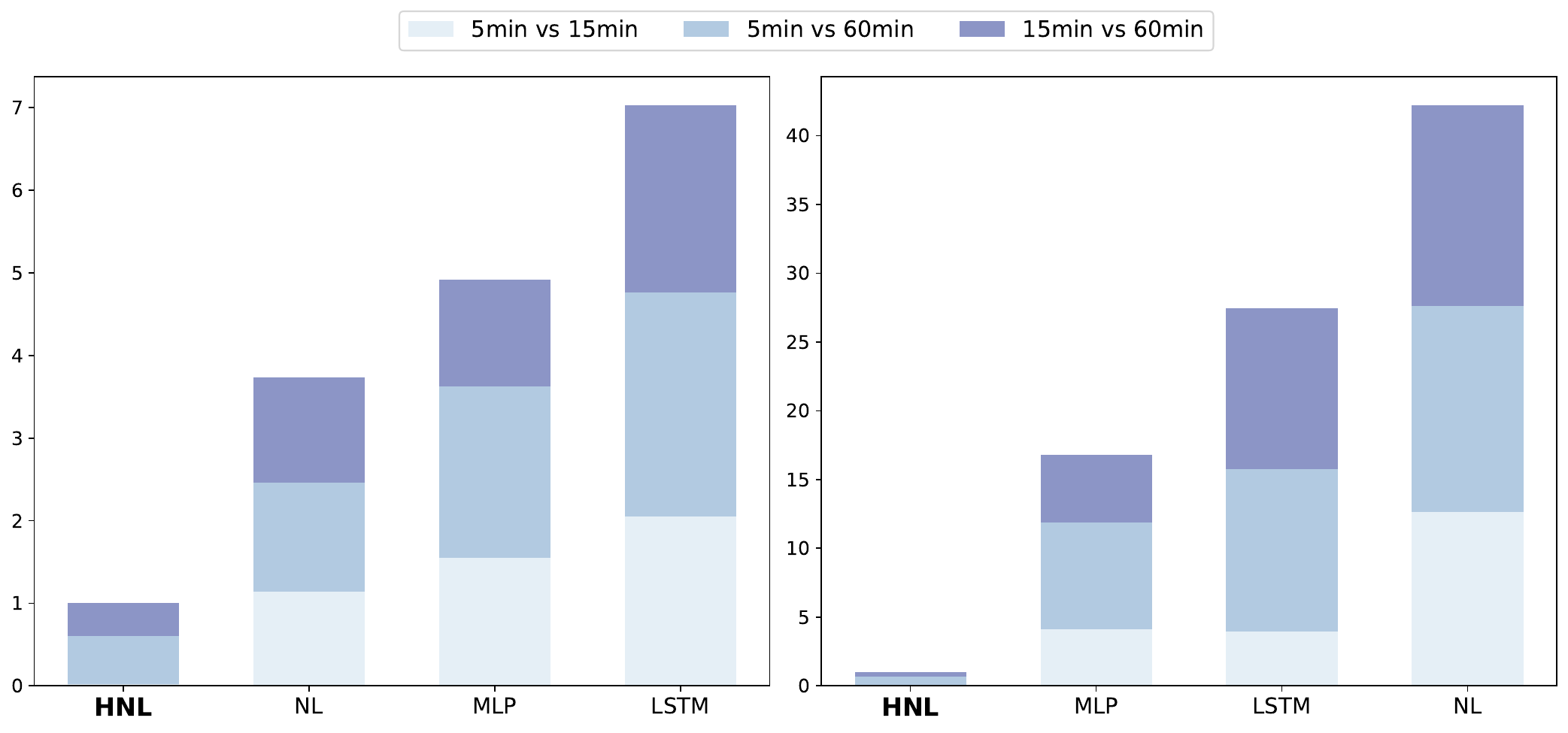}
    \caption{MCE comparison. The TCE is summed by the MCE between each resolution pair. The TCEs are scaled based on the performance of the HNL.}
    \label{fig:mce}
\end{figure}

Fig.~\ref{fig:mce} presents the comparison of our proposed HNL and uncoordinated models in terms of consistency. The most obvious inconsistency happens between the highest and the lowest resolution, i.e., 5-minute and 60-minute. The proposed HNL brings significant advantages in both electricity load and wind power forecasting. In contrast, the raw forecasts from the benchmarks without any coordination apparently cause extremely high consistency errors. 

\begin{figure}[htb]
    \centering
    \includegraphics[width=\textwidth]{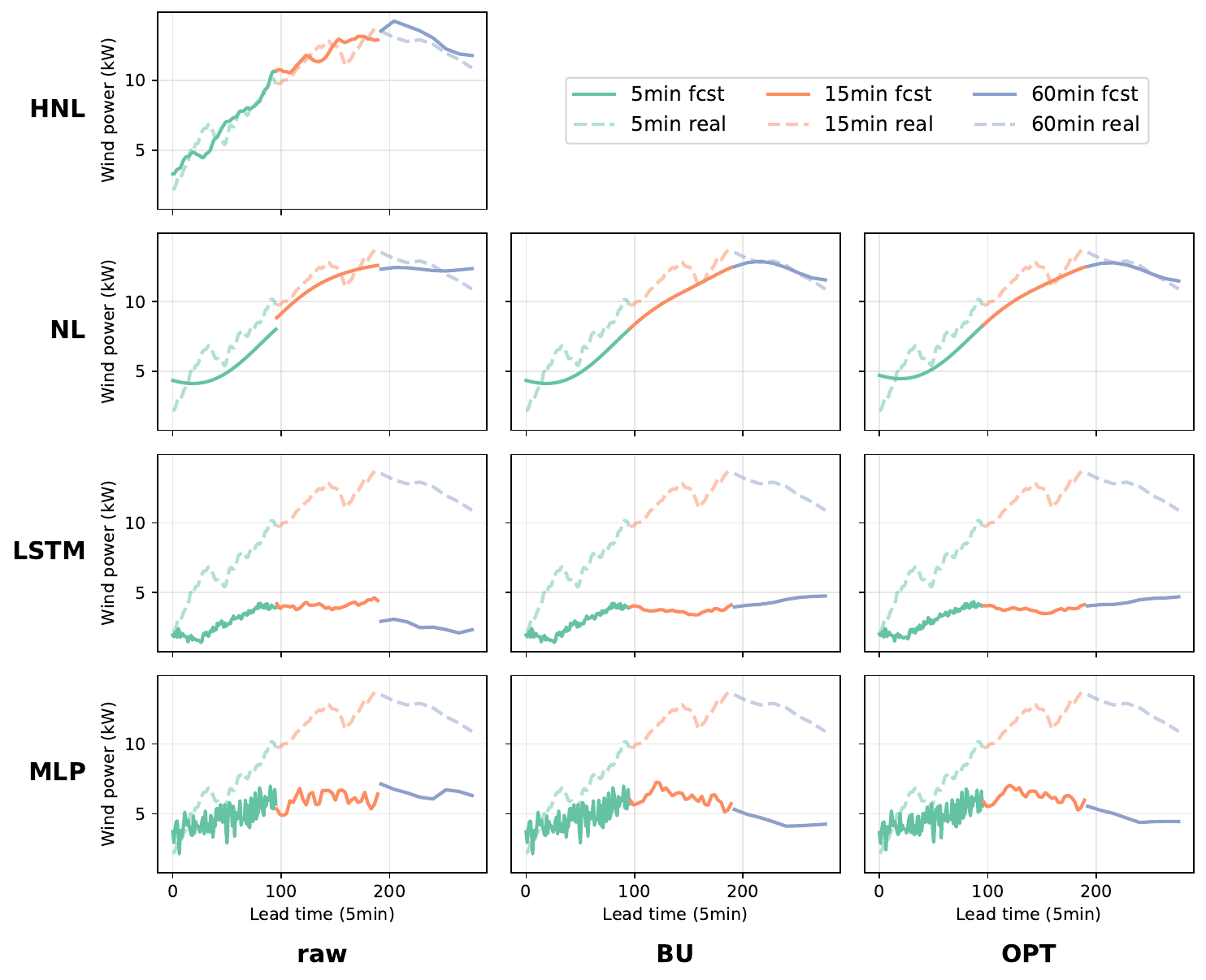}
    \caption{Multi-resolution forecasts comparison on wind power. One forecasting day is separated into three even horizons. Each row represents the model type, and each column represents the coordination strategy. The proposed HNL belongs to raw forecasting methods since it doesn't experience any coordination.}
    \label{fig:consistent}
\end{figure}

Besides, the leading position of HNL in wind power forecasting is more striking than electricity load forecasting. Our proposed HNL is at least 10 times better than the benchmarks in the wind power forecasting case. Fig.~\ref{fig:consistent} depicts an example of multi-resolution forecasts comparison for wind power generation. 

From the perspective of accuracy, traditional deep learning-based models, LSTM and MLP, fail to produce accurate forecasts in terms of both time and frequency domains. Especially in high-resolution situations, forecasts from MLP-related models introduce redundant high-frequency information. Though the forecasts from NL-related models track the trend of wind power, they fail to represent the information across relevant frequency ranges, owing to the limited frequency learning capability. These phenomena reflect the conclusion that the HNL method yield more accurate forecasts when assessed in both time and frequency domains. 

From the perspective of consistency, there appears abrupt gaps during the transition of forecasting resolutions among raw forecasts from benchmarks. For example, forecasts from LSTM somehow become lower if the forecasting resolution is switched from 15-minute to 60-minute.
Both post-coordination strategies (BU and OPT) eliminate these jumps in the raw forecasts, enhancing the consistency of the benchmark approaches. In contrast, even without any coordination strategies, the forecasts from HNL can still behave consistently across resolutions. The transition from one resolution to another is natural because of the framework design that generates energy forecasts at different resolutions hierarchically. Therefore, beyond the high accuracy in both time and frequency domains, forecasts from HNL are highly consistent across resolutions.

\subsection{Supportive effects for decision making}

The final goal of energy forecasting is to support decision-making, for a range of operational problems. Besides the accuracy and consistency of the forecasts, we also analyze the operational benefits brought by the forecasts in the downstream decision-making process. 

Specifically, we start with the day-ahead energy scheduling problem for a local region. It is a cost minimization problem that takes the 24-hour load forecast as the demand to be met, and outputs the optimal schedule of energy supply for each time step for the next day. Though the scheduling horizon is fixed to 24 hours, the scheduling resolution can vary, depending on how detailed the schedules are expected. Consequently, for different scheduling resolutions, the forecasts with the corresponding resolutions are required. Since it is impossible to forecast perfectly, additional real-time operations like external energy purchases are needed to balance the real energy demand and scheduled energy supply. This will cause extra operation costs. Therefore, regarding the extra costs, we evaluate the operational effects of the day-ahead forecasts at each resolution. The details of the settings of the day-ahead energy scheduling problem can be found in the supplementary documents.

Fig. \ref{fig:operation}(a)(b)(c) illustrate the comparison of operational costs from all forecasting models at different resolutions. We observe that with the resolution becoming higher, the advantage of our proposed HNL is more remarkable. Compared to the baseline model, the percentage savings of the operational costs are 14.8\%, 19.9\%, and 20.6\% at the resolutions of 60-minute, 15-minute, and 5-minute, respectively. For the 60-minute  resolution, the prevailing machine-learning model MLP even achieves lower operation costs. The possible reason is that hourly day-ahead forecasts only require 24 prediction points. The number of output dimensions is relatively small and contains limited frequency information. Traditional machine learning methods can perform well in such low-dimension forecasting settings. 


However, when it comes to the higher resolution cases, the energy curves will reveal more detailed frequency information. It may also affect the operation results. We collected each day-ahead forecast from all the models and computed the corresponding temporal RMSE, frequency RMSE, and total operational cost. Fig.~\ref{fig:operation} (d) depicts that forecasting errors in both time and frequency domains generally follow a linear relationship. Lower errors jointly in the time and frequency domains lead to lower additional operation costs. The blue shaded part shows a large proportion of special situations. The temporal errors of the forecasts are at a similar level, but the frequency errors varies a lot. In this vertical slice, we observe that with similar temporal accuracy, generally, the lower the errors in the frequency domain are, the lower the additional operational costs are in the day-ahead scheduling. Therefore, while our proposed HNL approach achieved similar temporal performance with the state-of-the-art benchmarks, the performance advantage of HNL in the frequency domain makes it substantially better in supporting decision-making.

\begin{figure}[htb]
\centering  
\subfigure[operational costs (5-minute)]{   
\begin{minipage}{0.31\textwidth}
\centering    
\includegraphics[width=\textwidth]{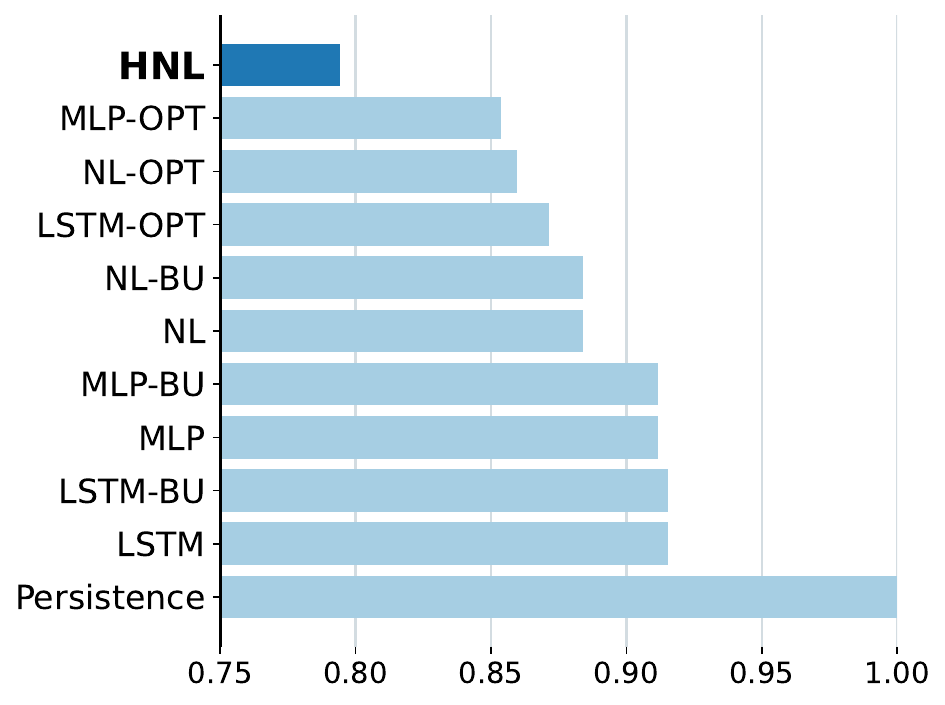}  
\end{minipage}
}
\subfigure[operational costs (15-minute)]{ 
\begin{minipage}{0.31\textwidth}
\centering    
\includegraphics[width=\textwidth]{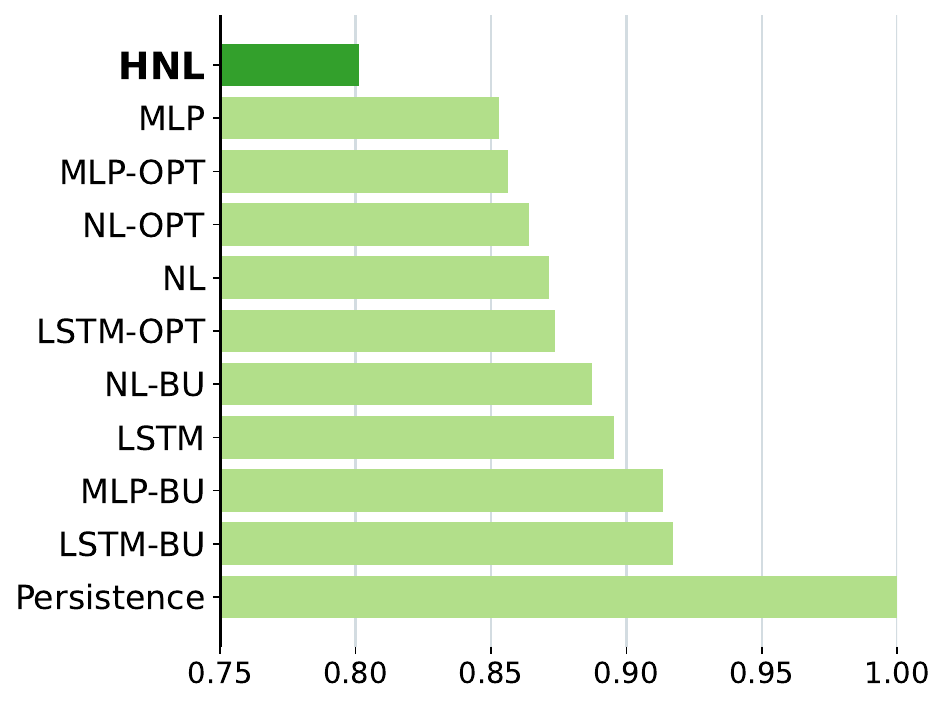}
\end{minipage}
}
\subfigure[operational costs (60-minute)]{   
\begin{minipage}{0.31\textwidth}
\centering    
\includegraphics[width=\textwidth]{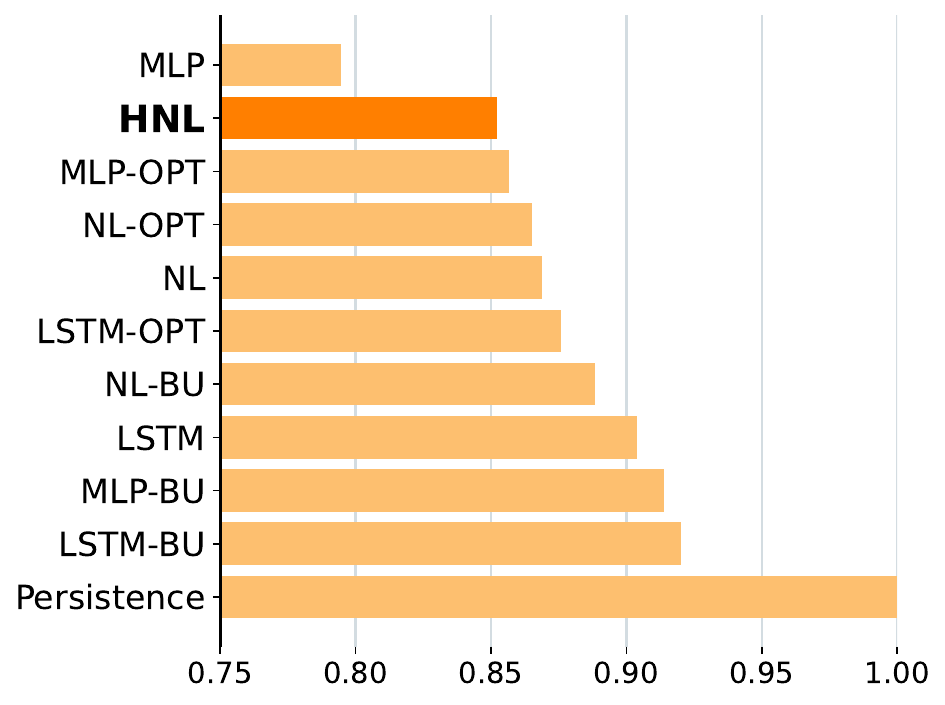}  
\end{minipage}
}\\
\subfigure[relationship of temporal error, frequency error and operational costs]{ 
\begin{minipage}{\textwidth}
\centering    
\includegraphics[width=\textwidth]{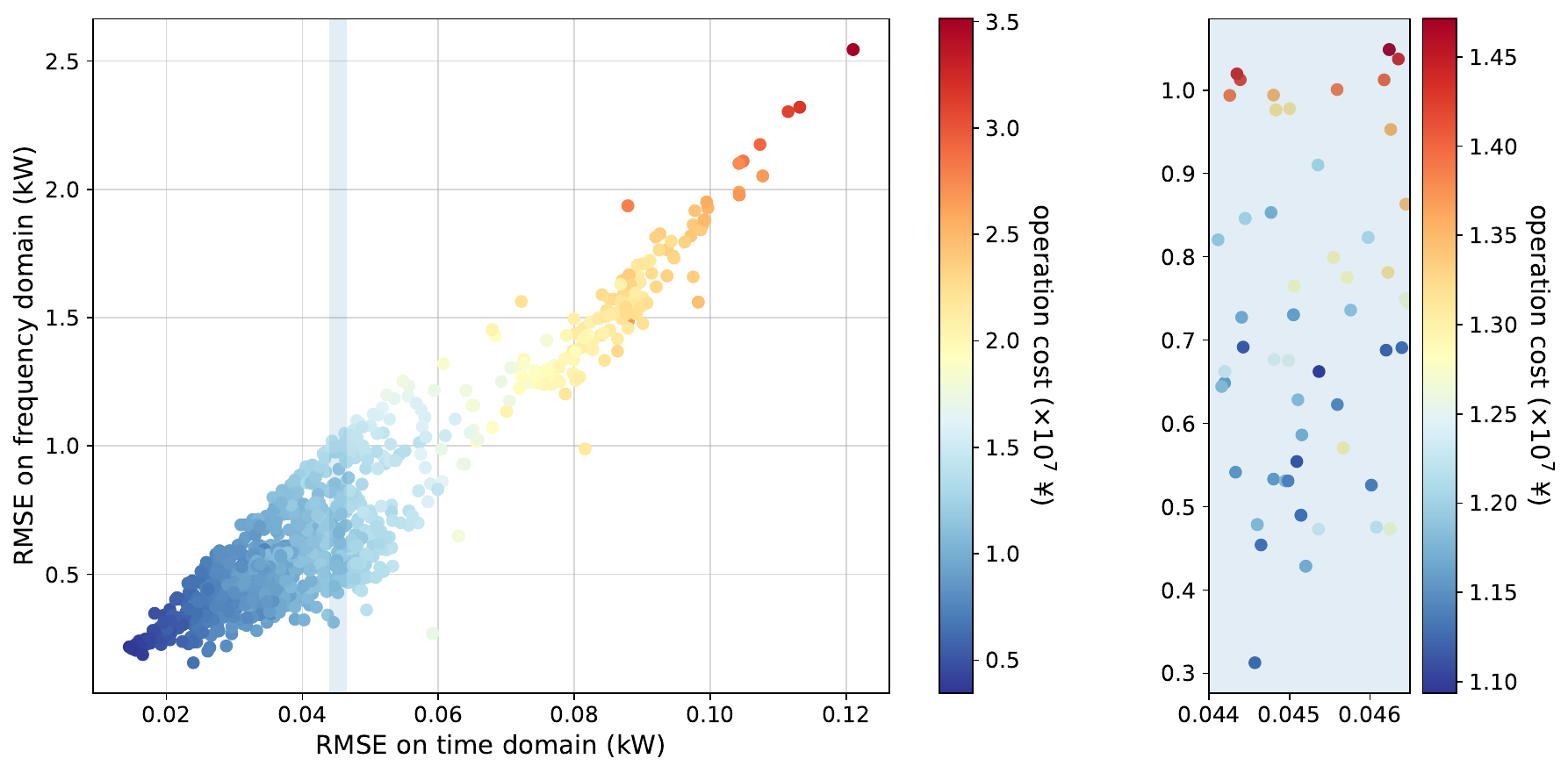}
\end{minipage}
}
\caption{Comparison of total operational costs in day-ahead scheduling. (a),(b) and (c) show the comparison based on 5-minute, 15-minute, and 60-minute  resolution forecasts, respectively. The operational costs in (a), (b), and (c) are scaled based on the performance of the Persistence model. (d) shows the relationship among temporal error, frequency error, and operational costs at the highest resoltuion.}    
\label{fig:operation}    
\end{figure}

In addition to day-ahead scheduling, energy system operators launch intra-day scheduling on a regular basis (for example, every 4 hours) to modify the day-ahead schedules. Distinct from day-ahead scheduling, intra-day scheduling concentrates on the operation decisions for a shorter horizon (usually 4 hours) and higher resolution (at least 15-minute). This is also a cost minimization problem which requires the updated energy forecasts at the according horizon and resolution.

Here, to conduct a more realistic case study, we jointly consider the day-ahead and intra-day scheduling, and denote as integrated scheduling. Apparently, in this case, multi-resolution energy forecasts cooperatively contribute to the final operation decisions. To quantify the value of multi-resolution forecasts in this context, we calculate the total operation costs, including the day-ahead costs for energy arrangements, the intra-day costs for adjustments, and the real-time costs for energy balancing. Details on integrated scheduling are attached in the supplementary documents.

\begin{figure}[htb]
    \centering
    \includegraphics[width=\textwidth]{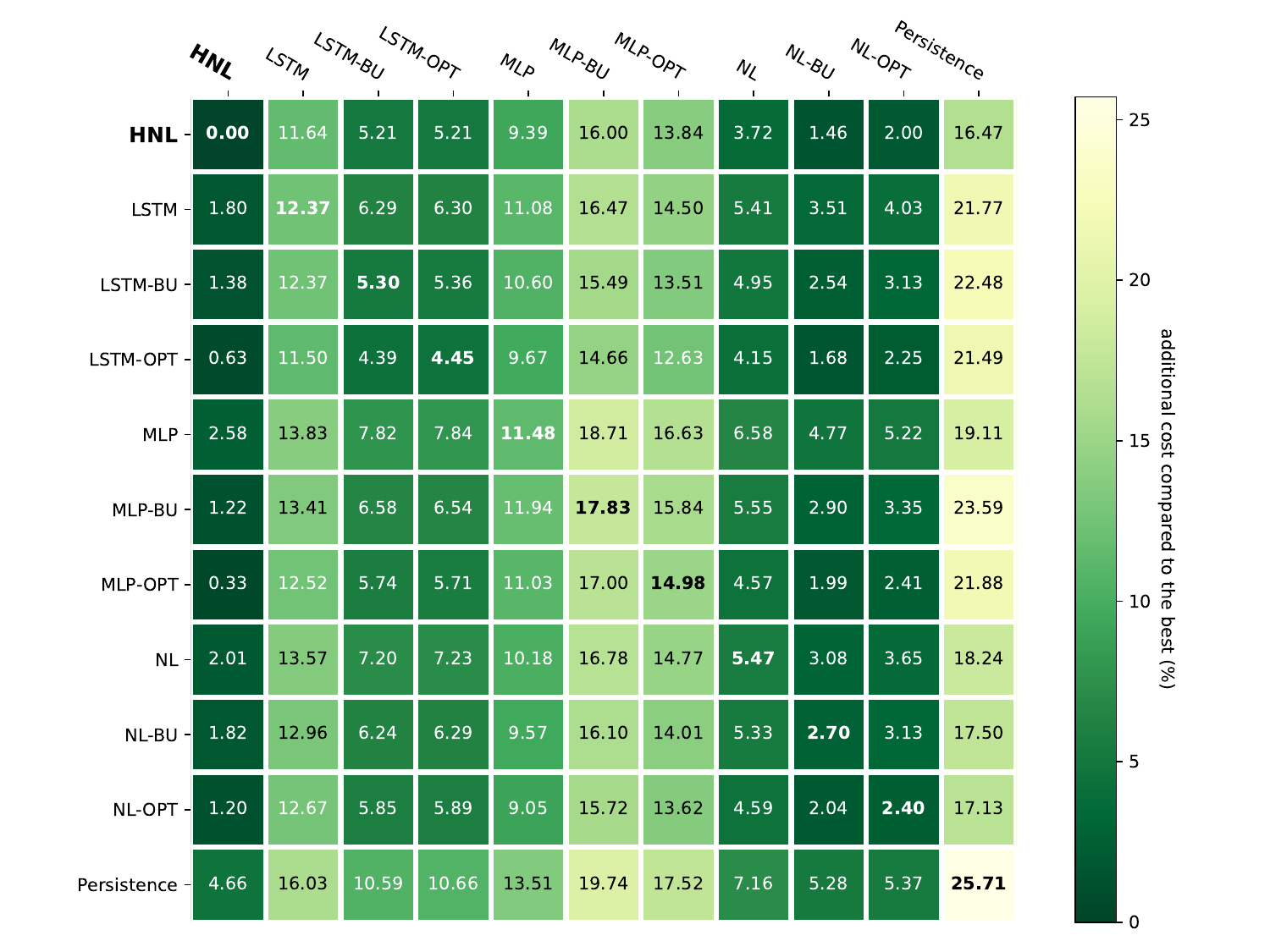}
    \caption{Comparison of operational costs for integrated scheduling.  Columns: wind forecasting methods. Rows: electricity load forecasting methods.  Entries: additional costs compared to the best in percentage value. Zero means the most economic situation.}
    \label{fig:loadwind}
\end{figure}

To fully illustrate the operational effects of both energy demand and supply forecasts, we enumerate all the possible pairs of forecasting approaches for electricity load and wind power generation, and calculate the resulting total operational costs. Fig.~\ref{fig:loadwind} presents the results in integrated scheduling of different model pairs. It is proved that our proposed HNL brings significant supportive effects in this large-scale realistic integrated scheduling case. Compared with the Persistence-based methods, the saved costs are more than 25\%. We also observe that, in this case, the differences between each column were remarkable, which means the wind forecasting methods dominate in the integrated scheduling. This is because the penetration rate of the wind power is set as a relatively high ratio (50\%). The results with other wind power penetration rates can be found in the supplementary documents.

Therefore, with high accuracy in both the time and frequency domain, our proposed energy forecasting framework, HNL, can provide more supportive forecasts for day-ahead scheduling. The relationship between temporal error, frequency error, and operation costs is explored as well. The striking advantage of HNL continues to be effective in the integrated scheduling process.

\section{Discussion}
In this paper, an innovative energy forecasting framework, HNL, is proposed for multi-resolution forecasting, providing unified modeling at different resolutions. Without the need for multiple models, the proposed framework is designed to characterize the hierarchical frequency information of the energy data based on the Laplace transform. Forecasts at desired resolutions can be flexibly produced by aggregating corresponding temporal components based on the Shannon sampling theorem.

Case studies have demonstrated that the HNL framework attains satisfactory accuracy improvement on the competitive benchmarks and baseline model. Thanks to the hierarchical framework design, more seamless forecasts across the resolutions can be generated than benchmarks. The focus on frequency learning gives the HNL an additional advantage in the accuracy in the frequency domain where the HNL can capture more precise variation patterns of the energy data. 
More importantly, in the follow-up day-ahead energy scheduling, this advantage of HNL can lead to better schedules with lower operation costs. The analysis of the relationship between temporal error, frequency error, and operation costs further implies that with similar temporal accuracy, forecasts with more precise frequency information may lead to better decisions. Consequently, it is suggested that accuracy in both time and frequency domains matters in energy forecasting, especially for the downstream decision-making. Besides, in the complicated integrated scheduling, multi-resolution forecasts of electricity load and wind power from HNL resulted in a salient reduction in terms of the total operation costs as well.

Though the HNL framework has achieved a satisfying performance on energy forecasting, there are still some aspects that deserve to be further explored. First, when it comes to a very high-resolution situation, there exists uncertain patterns in the energy curves, especially in the context of renewable energies \cite{wang2023inherent, kaack2017empirical}, which is the nature of the energy data. Therefore, it is necessary to extend the current HNL framework into a probabilistic one to model the uncertainty of the forecasts. 
Besides, Laplace Transform can only tell what kind of temporal components exist in the data but can not tell when and what components are dominant. We will further inspect how to treat other time-frequency transformations as decoders in energy forecasting. In this way, we can analyze the forecast results and their operational effects for specific time periods (for example, peak hours) in a refined manner.

Nevertheless, the proposed framework obtains fair performance over competitive benchmarks, which provides a unified perspective to deal with multi-resolution energy forecasting by capturing the hierarchical frequency information in the energy data. In addition, forecasts from the proposed framework are proven to have prominent supportive effects for energy scheduling, which fulfill the research gap of multi-resolution energy forecasting and its resulting operation effects.

\section{Method}\label{chap:method}
\subsection{Analysis of Fourier-based ILT}
As mentioned in Section 2, Laplace decoders are the core of our framework, relying on the Laplace transform and ILT. Given a  temporal function $f(t)$ of energy data, the corresponding Laplace function $\bar{f}(s)$ through Laplace transform is defined as:
\begin{equation}
    \mathscr{L}\{f(t)\}=\bar{f}(s)=\int_0^{\infty} e^{-s t} f(t) d t. 
\end{equation}
Similarly, given a Laplace function $\bar{f}(s)$, the temporal function can be inversely transformed as its definition shows:
\begin{equation}\label{eq:ilt}
\mathscr{L}^{-1}\{\bar{f}(s)\}=f(t)=\frac{1}{2 \pi i} \int_{\gamma-i \infty}^{\gamma+i \infty} e^{s t} \bar{f}(s) d s.
\end{equation}

According to \cite{crump1976numerical}, Eq.~\eqref{eq:ilt} can be manipulated into another form and discretized through the trapezoid rule: 
\begin{align}
    f(t) & = \frac{e^{\gamma t}}{\pi} \int_0^{\infty}  \Re \left(\bar{f}(s) e^{i\omega t}\right) d \omega \\
    & \approx \frac{e^{\gamma t}}{T} \left[\frac{\bar{f}(\gamma)}{2}+\sum_{k=1}^{N} \Re \left\{\bar{f}\left(\gamma+\frac{i k \pi}{T}\right) \exp \left(\frac{i k \pi t}{T}\right)\right\}\right],\label{eq:dis}
\end{align}
where $T$ and $\gamma$ are usually set as constants, and $N$ controls the number of discrete integral terms, dominating the discretization of ILT.

The idea of Laplace decoders \cite{holt2022neural} is utilizing a neural network $g_{\omega}$ to learn the Laplace function inside the data, i.e.
\begin{equation}\label{eq:ld}
\mathbf{\bar{f}}(\mathbf{s}) = g_{\omega}(\mathbf{s}, \mathbf{h}), \mathbf{s} =  [s_0, s_1, \cdots, s_N]^\top, s_k=\gamma+\frac{i k \pi}{T} \end{equation}
where $\mathbf{\bar{f}}(\mathbf{s})=[\bar{f}(s_0), \bar{f}(s_1),\cdots,\bar{f}(s_N)]^\top$; $\omega$ is the trainable weights in the neural network; $\mathbf{h}$ is the hidden states extracted from the input data through the encoder. Then, given the learned $\mathbf{\bar{f}}(\mathbf{s})$, we can obtain the inversely transformed temporal function $f(t)$ through Eq.~\eqref{eq:dis}. With the temporal function $f(t)$, we can forecast by inferring the time steps, for example $f(t_n)$.

More importantly, if we further rewrite the Eq.~\eqref{eq:dis} with the Eular formula, then the following form can be obtained:
\begin{align}
    f(t)
    & = \frac{e^{\gamma t}}{T} \left[\frac{\bar{f}(s_0)}{2}+\sum_{k=1}^{N} \Re \left\{\bar{f}\left(s_k\right) \left(\cos(\frac{k\pi t}{T})+i\sin(\frac{k\pi t}{T})\right)\right\}\right].
\end{align}
From the above equation, we observe that the inversely transformed $f(t)$ explicitly consists of cosine waves with exponential effects at different frequencies ($f_c^k=k/2T$). The maximum frequency contained ($f_c^{max}=N/2T$) is apparently controlled by the parameter $N$. This implies that if we adopt Laplace decoders as the output layers in the framework, the frequencies contained in the forecasts can be analytically derived.

\subsection{Connection of frequencies and resolutions}
Since the ILT makes the frequency information transparent in the forecasts, this subsection aims to illustrate the connection of frequencies and the resolutions of energy data.

An energy time series basically consists of discrete samples from the continuous temporal domain. Assume that a energy time series is recorded on a time range $[0,...,t_n]$ with $n$ samples,  the resolution of this time series is defined as $f_r = {n}/{t_{n}}$.

According to Nyquist–Shannon sampling theorem \cite{shannon1949communication}, for a given sampling resolution $f_r$, the maximum informative frequency that can be expressed in the original data is no more than $1/2 f_r$, i.e., $f_c^{max}\leq 1/2 f_r$. 
Therefore, if we want to reconstruct the temporal function through \eqref{eq:dis} without information loss, then the threshold $N$ can be set as:
\begin{equation}\label{eq:bign}
   N = 2 T \cdot 1/2 f_r = T f_r. 
\end{equation}
It tells us that by setting $N=Tf_r$, we can include all the necessary information at the resolution $f_r$. Also, if we want to reconstruct at a lower resolution, for instance $f_r^\prime$, with precise frequency information, then we can just lower $N$ to $N^\prime = Tf_r^\prime$. Therefore, the parameter $N$ acts like a frequency filter, and the connection of frequencies and resolutions can be analytically expressed in Eq.~\eqref{eq:bign}. A toy example of the effects of frequency parameter is included in the supplementary document.

\subsection{Multiple Laplace decoders}
Now assuming there are $m$ ascending sampling resolutions that we are interested in and are not beyond the original sampling resolution, i.e. $f_r^1,\cdots,f_r^m$ and $f_r^m \leq f_r$, we can then set the corresponding frequency parameter for each resolution according to Eq.~\eqref{eq:bign}, resulting in $N_1, \cdots, N_m$ and $N_m \leq N$. These frequency parameters will act as anchors in Eq.~\eqref{eq:dis}:
\begin{align}
    f_m(t)
   &= \frac{e^{\gamma t}}{T} \bigg[ \underbrace{\frac{\bar{f}(s_0)}{2}+\sum_{k=1}^{N_1} S_k(t)}_\text{$TC_1$}  + \underbrace{\sum_{k=N_1+1}^{N_2} S_k(t)}_\text{$TC_2$}+\cdots+\underbrace{\sum_{k=N_{m-1}+1}^{N_m} S_k(t)}_\text{$TC_m$} \bigg],
\end{align}
where we denote $S_k(t)=\Re \left\{\bar{f}\left(s_k\right) \left(\exp \left(\frac{i k \pi t}{T}\right)\right)\right\}$ for simplicity and $TC_i$ represents the $i$th temporal component. 

\begin{figure}[htb]
    \centering
    \includegraphics[width=\textwidth]{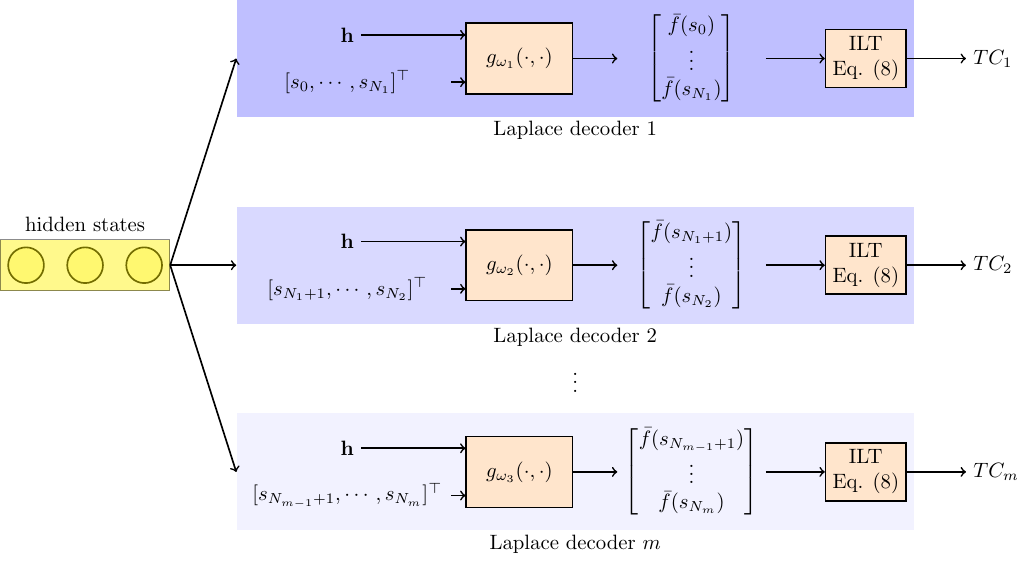}
    \caption{The details of multiple Laplace decoders}
    \label{fig:frame_detail}
\end{figure}

Theoretically, if we train a Laplace decoder, as illustrated in Eq.~\eqref{eq:ld}, with the output Laplace function values $\mathbf{\bar{f}}(\mathbf{s})$, we can then generate forecasts at any desired resolution $f_r^i \in \{f_r^1,\cdots,f_r^m\}$ with full frequency information by collecting the corresponding groups of temporal components $TC_1, \cdots, TC_i$.
However, in practice, the energy data at a high resolution will apparently lead to a huge frequency parameter $N$. This can cause the curse of dimensionality during the training of this large Laplace decoder. The failure of training a large Laplace decoder is demonstrated in the supplementary document. To this end, as illustrated in Fig.~\ref{fig:frame_detail}, we proposed to adopt multiple compact Laplace decoders to separately learn the Laplace functions on the interested frequency bands:
\begin{equation*}
    \begin{cases}
        \mathbf{\bar{f}}(\mathbf{s}_1) = g_{\omega_{1}}(\mathbf{s}_1, \mathbf{h}), & \mathbf{s}_1 =  [s_0, s_1, \cdots, s_{N_{1}}]^\top \\
    \mathbf{\bar{f}}(\mathbf{s}_2) = g_{\omega_{2}}(\mathbf{s}_2, \mathbf{h}), & \mathbf{s}_2 =  [s_{N_{1}+1}, s_1, \cdots, s_{N_{2}}]^\top,  \\
    &\vdots\\
    \mathbf{\bar{f}}(\mathbf{s}_m) = g_{\omega_{m}}(\mathbf{s}_m, \mathbf{h}), & \mathbf{s}_m =  [s_{N_{m-1}+1}, s_1, \cdots, s_{m}]^\top, 
    \end{cases}
\end{equation*}

In summary, through the analysis of Fourier-based ILT, we can fully understand the energy forecasts produced from ILT by explicitly decomposing them into cosine waves (with exponential effects) at different frequencies. Then, utilizing the Shannon sampling theorem, the frequencies in the forecasts and their resolutions of them are bridged analytically. The desired resolutions set their corresponding anchors in the frequency domain, splitting the whole domain into frequency bands. Multiple Laplace decoders are subsequently configured on these frequency bands and trained to learn the respective Laplace functions. After the training stage, given a desired resolution, we can assemble the forecasts by aggregating the related temporal components with ease.

\subsection{Methods for comparison}
In terms of forecasting without post-coordination, we include Persistence, MLP, LSTM and NL as competing methods. The former three are well-known benchmarks in energy forecasting \cite{hong2020energy}. Therefore, here, we give some details on NL and the two post-coordination strategies, i.e., BU and OPT.

Compared to our framework with hierarchical Laplace decoders, NL only adopts one single Laplace decoder to learn the Laplace function. However, as mentioned before, the curse of dimensionality happens if we set the frequency parameter $N$ as a huge integer. The failure of NL with a large $N$ on forecasting can be found in the supplementary document. Therefore, we have to apply a relatively small $N$ in NL and train it at each resolution, which causes the loss of frequency information and redundant training. This also reflects the effectiveness and efficiency of our proposed HNL.

As for the coordination strategies, the BU strategy means that we only forecast on the highest resolution, and the forecasts at the lower resolutions are all generated by downsampling. OPT strategy makes full use of forecasts at all desired resolutions, solving a generalized least squares problem with the forecasts at resolutions and the temporal structure as inputs \cite{athanasopoulos2017forecasting}. 

\subsection{Performance metrics}
The performance metrics used in Section 2 for evaluating the energy forecasts include RMSE in the time domain, RMSE in the frequency domain, TCE, and the total operation costs in the energy scheduling. 

First, we assume that given the information on and before time $t$, the $k$th forecasts in the test set $D_{\text{test}}$ at the resolution $f_r^i$ on the following steps $n_i$ are denoted as $\mathbf{\hat{y}}_{i,k} = [\hat{y}_{t+1|t}^{k}, \cdots, \hat{y}_{t+n_i|t}^{k}]$. The corresponding real energy data at this resolution is denoted as $\mathbf{y}_{i,k} = [{y}_{t+1}^{k}, \cdots, {y}_{t+n_i}^{k}]$. Given the fixed forecasting horizon (24 hours in this paper), the forecasting steps $n_i$ change according to the desired resolution $f_r^i$. Therefore, RMSE in the time domain for this $j$th forecast at the resolution $f_r^i$ and the overall average RMSE on the whole test set can be respectively calculated as:
\begin{align}
    RMSE_{\text{time}}^{i, k} &= \sqrt{\frac{1}{n_i}\sum_{n=1}^{n_i}(y_{t+k}^{k}-\hat{y}_{t+n|t}^{k})^2},\\
    RMSE_{\text{time}}^{i} &= \frac{1}{|D_{\text{test}}|}\sum_{k=1}^{|D_{\text{test}}|}RMSE_{\text{time}}^{i, k},
\end{align}
where $|D_{\text{test}}|$ denotes the size of the test set. In this way, the average RMSE at the resolution $f_r^i$ over the whole test set can be calculated.

Then, we perform fast Fourier transform \cite{brigham1988fast} (FFT) on the $k$th real energy data and the predicted one at the highest desired resolution $f_r^{m}$, and obtain the amplitudes of frequencies respectively, i.e. $\mathbf{y}_{m,k}^{freq} = [{y}_{f_1}^{k}, \cdots, {y}_{f_{M}}^{k}]$ and $\mathbf{\hat{y}}_{m,k}^{freq} = [\hat{y}_{f_1}^{k}, \cdots, \hat{y}_{f_{M}}^{k}]$. Here, we denote the total number of frequencies demonstrated in FFT as $M$. The real amplitude ${y}_{f_{k}}^{k}$ and the predicted amplitude $\hat{y}_{f_{k}}^{k}$ are shown at the frequency $f_k$. In this way, we can calculate the RMSE in the frequency domain as well:
\begin{align}
    RMSE_{\text{freq}}^{m, k} &= \sqrt{\frac{1}{M}\sum_{n=1}^{M}({y}_{f_1}^{k}-\hat{y}_{f_n}^{k})^2},\\
    RMSE_{\text{freq}}^{m} &= \frac{1}{|D_{\text{test}}|}\sum_{k=1}^{|D_{\text{test}}|}RMSE_{\text{freq}}^{m, k}.
\end{align}
In this way, in the frequency domain, the average RMSE at the highest resolution $f_r^m$ over the whole test set can be calculated. The unit of both RMSEs in both time and frequency domain is the same as the original energy data, i.e. kW in our case studies.




In terms of evaluating consistency, the $j$the MCE in the test set between two forecasts at different resolutions $f_r^i$ and $f_r^j$ are defined as:
\begin{equation}
    MCE(\mathbf{\hat{y}}_{i,k}, \mathbf{\hat{y}}_{j,k}) = \frac{1}{n}\left\Vert ds(\mathbf{\hat{y}}_{j,k}, f_r^i)  - \mathbf{\hat{y}}_{i,k} \right\Vert_{2}^{2},
\end{equation}
where the function $ds(\cdot)$ takes in the forecasts at resolution $f_r^j > f_r^i$ and downsamples them into the resolution $f_r^i$. Then, the Euclidean distance of the downsampled forecasts and the original forecasts at the resolution $f_r^i$ will be calculated. For each resolution pair $(f_r^i, f_r^j)$, we will calculate as above and sum up to obtain the final total consistency error (TCE) of the $k$th forecasts:
\begin{equation}
    TCE_k = \sum_{i=1}^{m} \sum_{j=i+1}^{m} MCE(\mathbf{\hat{y}}_{i,k}, \mathbf{\hat{y}}_{j,k}).
\end{equation}
We also calculate the average TCE over the whole test set and have $TCE = \frac{1}{|D_{\text{test}}|}\sum_{k=1}^{|D_{\text{test}}|}TCE_{k}.$

As for energy scheduling, we consider an economic dispatch problem in both day-ahead and intra-day fashion, including the ramping constraints of generators, battery storage devices and so on. The details of the specific optimization problems can be found in the supplementary document.

\section{Data availability}
The load and wind power datasets are available respectively at \url{https://dataverse.harvard.edu/dataset.xhtml?persistentId=doi:10.7910/DVN/X9MIDJ} and \url{https://www.nrel.gov/grid/wind-toolkit.html}. 
The NWP data is available from the ECMWF website \url{https://www.ecmwf.int/en/forecasts/dataset/operational-archive} after registration as real-name users (for researchers based in Europe, at least). The above mentioned datasets have been processed and deposited in our GitHub repository \url{https://github.com/WillWang1113/MultiresolutionForecasting}.

\section{Code availability}
The code and data for the experiments are publicly available on \url{https://github.com/WillWang1113/MultiresolutionForecasting}.

\bibliographystyle{unsrt}
\bibliography{ref}

\begin{appendices}

\section{Energy schedulling}
\subsection{Day-ahead scheduling}

\begin{align}
 \min _{\substack{\{P_{j i}, P^c_i,\\P^d_i,SOC_i, \phi_i \}}}&   C_g + C_b\\
s.t. \quad & C_g = \sum_{i=1}^{N} \sum_{j=1}^J (a_j P_{j i}^2+b_j P_{j i}+c_j) \label{eq:cg} \\
& C_b = \sum_{i=1}^{N} \pi \cdot (\eta_c P^c_i + P^d_i / \eta_d) \label{eq:cb} \\
& \sum_{j=1}^J P_{j i} + P^d_i=\widehat{L}_{i} + P^c_i  \quad: \forall i, \label{eq:balance}\\
& 0 \leq P_{j i} \leq \overline{P_j} \quad: \forall j, \forall i, \label{eq:ramp1}\\
& P_{j i}-P_{j(i-1)} \leq R_j^U \cdot \Delta t \quad: \forall j, \forall i \geq 2, \\
& P_{j(i-1)}-P_{j i} \leq R_j^D \cdot \Delta t \quad: \forall j, \forall i \geq 2, \label{eq:ramp2}\\
& Cap \cdot SOC_i = Cap \cdot SOC_{\text{init}} +(P^c_{i}\cdot \eta_c - P^d_{i} / \eta_d)\cdot \Delta t \quad:  i=1, \label{eq:bat1}\\
& Cap \cdot SOC_i = Cap \cdot SOC_{i-1} + (P^c_{i}\cdot \eta_c - P^d_{i} / \eta_d)\cdot \Delta t \quad:  \forall i \geq 2, \\
& \underline{SOC} \leq SOC_{i} \leq \overline{SOC} \quad: \forall i, \\
& 0 \leq P^c_{i} \leq {P^{\text{batt}}_j} \cdot \phi_i \quad: \forall i, \\
& 0 \leq P^d_{i} \leq {P^{\text{batt}}_j} \cdot (1-\phi_i) \quad: \forall i, \\
& \phi_i \in \{0, 1\}, \quad: \forall i.\label{eq:bat2}
\end{align}

In the day-ahead scheduling problem, on the basis of power balance (Eq.~\eqref{eq:balance}), we consider the generators with ramping constraints (Eq.~\eqref{eq:ramp1} - Eq.~\eqref{eq:ramp2}) for the basic supply and the battery storage (Eq.~\eqref{eq:bat1} - Eq.~\eqref{eq:bat2}) for flexible supply. $i$ is the index for time unit and $j$ is the index for the generators. $P_{j i}$ represents the power output of the $j$th generator at $i$th time unit. $P^c_{i}$ and  $P^d_{i}$ represent the charging and discharging power at $i$th time unit. $\phi_{i}$ is a binary variable which controls charging or discharging. $SOC_i$ is the state of charging of the battery. The power flow constraints are omitted for simplicity.

Given the load forecasts $\widehat{\mathbf{L}}=[\widehat{L}_{1},\cdots,\widehat{L}_{N}]$, the aim of the day-ahead scheduling is to give the day-ahead schedule $\mathbf{P}$ that minimizes the sum of the costs of the generators $C_g$ and the degradation costs of the battery $C_b$.  Further, we denote this day-ahead operation costs and the day-ahead schedule as a function of forecasts, i.e. $C_{da}(\widehat{\mathbf{L}}), \mathbf{P}(\widehat{\mathbf{L}})$ , and the operation costs based on the perfect forecasts as $C^{*}_{da} = C_{da}({\mathbf{L}}), \mathbf{P}^* = \mathbf{P}({\mathbf{L}})$.

Based on the day-ahead schedule of generators $P^{da}_{ji}$, the battery storage schedule will be re-adjusted to balance the supply and demand according to the real load $L$. This forms the following problem: 
\begin{align}
 \min _{\substack{\{P^c_i,P^d_i, SOC_i, \phi_i \}}}&   C_g + C_b\\
s.t. \quad & \eqref{eq:cg}-\eqref{eq:cb} \\
& \sum_{j=1}^J P_{j i} + P^d_i={L}_{i} + P^c_i \quad: \forall i, \\
& P_{j i}=P^{da}_{ji} \quad: \forall j, \forall i, \label{eq:same} \\
& \eqref{eq:bat1}-\eqref{eq:bat2}
\end{align}

Therefore, we can see that the costs of generators $C_g$ is not changing (Eq.~\eqref{eq:same}) while the costs of the battery $C_b$ will be re-calculated according to the difference between the scheduled supply and the real load. We also denote this real-time operation costs of day-ahead schedule as a function of day-ahead schedules $C_{rt}(\mathbf{P}(\widehat{\mathbf{L}}))$ and for the perfect forecasts, the corresponding cost is $C^*_{rt} = C_{rt}(\mathbf{P}^*)$. Consequently, in the perfect forecasting situation, we don't have to re-adjust the battery schedule and have $C^*_{rt} = C^*_{da}$. However, for the imperfect forecasts, we will have $C_{rt}(\mathbf{P}(\widehat{\mathbf{L}})) > C_{da}(\widehat{\mathbf{L}})$. To evaluate the operational effects of the forecasts caused by imperfection, we calculate the additional operation costs compared to the perfect situation and obtain $C_{rt} - C^{*}_{da}$.

\subsection{Integrated scheduling}
$$
\begin{aligned}
 \min _{P_{j i}, W_{i} Wc_{i}}& (C_g + C_w)\cdot \Delta t\\
s.t. \quad & C_w = WC \cdot W^c_{i}, \\
& W_{i}+\sum_{j=1}^J P_{j i}=\widehat{L}^{low}_{i} \quad: \forall i,\\
&  W_{i}+W^c_{i}=\widehat{W}^{low}_{i} \quad: \forall i,\\
& 0 \leq W_{i} \leq \widehat{W}^{low}_{i} \quad: \forall i, \\
& 0 \leq W^c_{i} \leq \widehat{W}^{low}_{i} \quad: \forall i, \\
& \eqref{eq:cg}, \eqref{eq:ramp1} - \eqref{eq:ramp2} \\
\end{aligned}
$$
In the large-scale integrated scheduling, we first considered a day-ahead scheduling problem as illustrated above. Here, considering the difference of scheduling resolution between day-ahead and intra-day scheduling, the objective function is scaled by the time interval.

We also include the wind power into the consideration. $W_i$ represents the wind power that is used as power supply while $W^c_i$ represents the wind power that is curtailed. The curtailed wind power will be punished as $C_w$. Beyond the load forecasts $\widehat{L}^{low}_{i}$, the wind forecasts $\widehat{W}^{low}_{i}$ are also needed. Both of them are required at a relatively low resolution for day-ahead scheduling, i.e. 60min level. The aim is to determine the schedules of the generators $P^{da}_{ji}$.
$$
\begin{aligned}
 \min _{\substack{\{W_i, W^c_i, P^c_i,\\P^d_i, SOC_i, \phi_i \}}}& (C_g + C_w + C_b) \cdot \Delta t \\
s.t. \quad & W_{i}+P^c_i+\sum_{j=1}^J P_{j i}=\widehat{L}^{high}_{i} +P^d_i \quad: \forall i,\\
& P_{ji} = P^{da}_{ji} \quad: \forall i,\forall j, \\
&  W_{i}+W^c_{i}=\widehat{W}^{high}_{i} \quad: \forall i,\\
& 0 \leq W_{i} \leq \widehat{W}^{high}_{i} \quad: \forall i, \\
& 0 \leq W^c_{i} \leq \widehat{W}^{high}_{i} \quad: \forall i, \\
& \eqref{eq:bat1} - \eqref{eq:bat2}.
\end{aligned}
$$

On the basis of the day-ahead schedule of generators $P^{da}_{ji}$, the battery storage schedules $P^c_i,P^d_i$ will be considered and determined to balance the supply and demand according to the more timely load and wind power forecasts $\widehat{L}^{high}$ and $\widehat{W}^{high}$. This intra-day scheduling will be run every 4 hours at a high resolution, i.e. 5min level, for the next 4 hours.

$$
\begin{aligned}
 \min _{\substack{\{W_i, W^c_i, P^p_i,P^n_i \}}}& (C_g + C_w + C_b + \sum_{i=1}^{N} (P^p_i \cdot \pi_p + P^n_i \cdot \pi_n)) \cdot \Delta t \\
s.t. \quad & W_{i}+P^c_i+P^p_i+\sum_{j=1}^J P_{j i}={L}^{high}_{i} +P^d_i+P^n_i \quad: \forall i,\\
& P_{ji} = P^{da}_{ji} \quad: \forall i,\forall j, \\
& P^c_i = P^{id,c}_{i} \quad: \forall i, \\
& P^d_i = P^{id,d}_{i} \quad: \forall i, \\
&  W_{i}+W^c_{i}={W}^{high}_{i} \quad: \forall i,\\
& 0 \leq W_{i} \leq {W}^{high}_{i} \quad: \forall i, \\
& 0 \leq W^c_{i} \leq {W}^{high}_{i} \quad: \forall i, \\
& \eqref{eq:bat1} - \eqref{eq:bat2}.
\end{aligned}
$$

Finally, given the schedules of generators $P^{da}_{ji}$, the battery $P^{id,c}_{i},P^{id,d}_{i}$ and the real-time load and wind power ${L}^{high}, {W}^{high}$, the real-time positive/negative imbalances $P^p_i, P^n_i$ will be unveiled, and the corresponding operations that compensate the imbalances will cause extra costs $P^p_i \cdot \pi_p + P^n_i \cdot \pi_n$. Consequently, the total costs of the integrated scheduling of this day can be obtained as the objective function shows in the above problem.

\section{Experimental results}
\subsection{Toy example}

Here, a toy example is shown to demonstrate the method to control the frequency we want to learn through Laplace decoder. The following figure shows the effect of frequency parameter $N$.
\begin{figure}[htb]
\centering  
\subfigure[Time domain ($N=33$)]{   
\begin{minipage}{0.476\textwidth}
\centering    
\includegraphics[width=\textwidth]{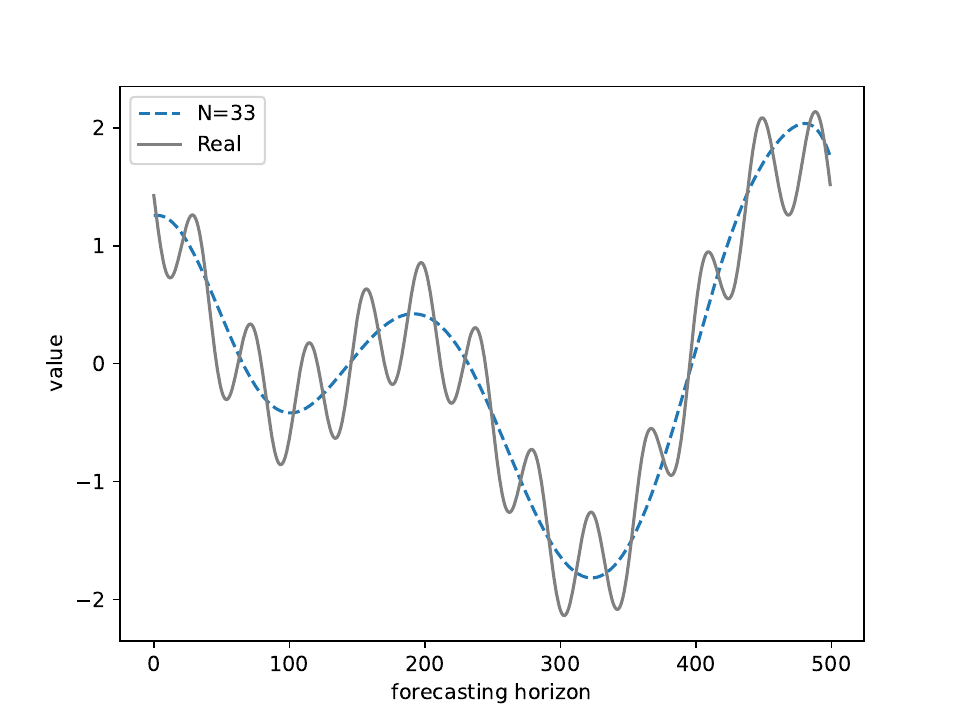}  
\end{minipage}
}
\subfigure[Time domain ($N=161$)]{ 
\begin{minipage}{0.476\textwidth}
\centering    
\includegraphics[width=\textwidth]{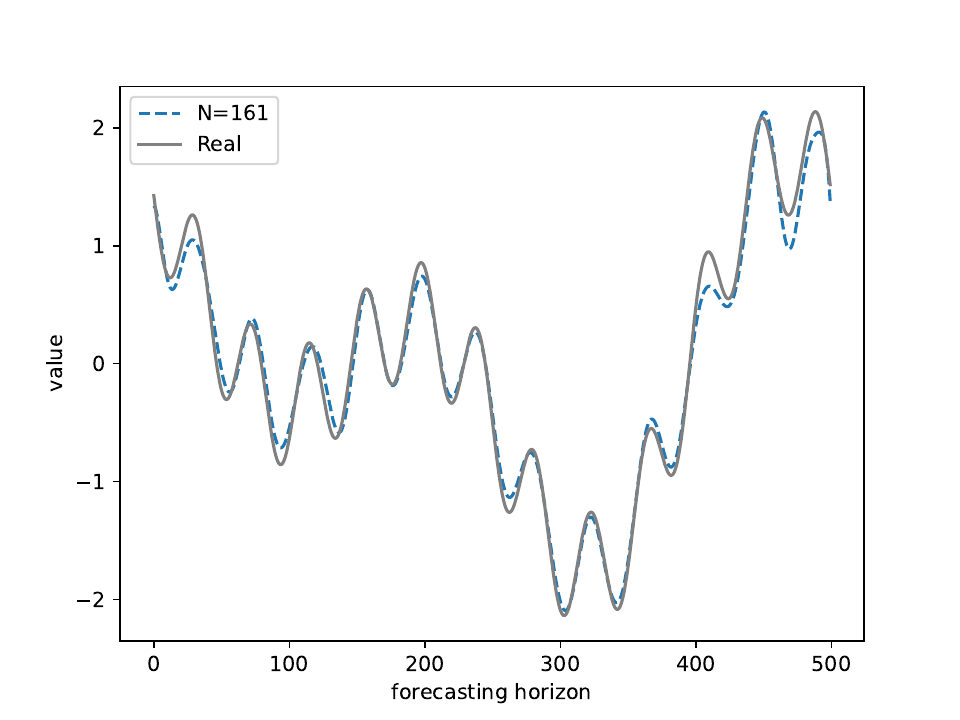}
\end{minipage}
}\\
\subfigure[Frequency domain ($N=33$)]{   
\begin{minipage}{0.476\textwidth}
\centering    
\includegraphics[width=\textwidth]{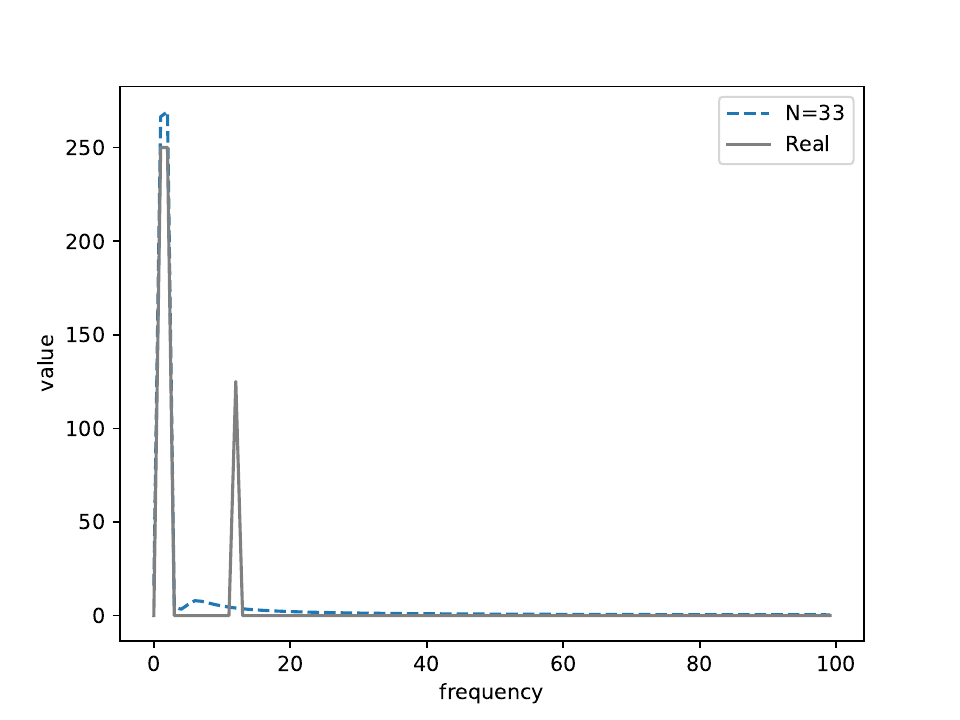}  
\end{minipage}
}
\subfigure[Frequency domain ($N=161$)]{ 
\begin{minipage}{0.476\textwidth}
\centering    
\includegraphics[width=\textwidth]{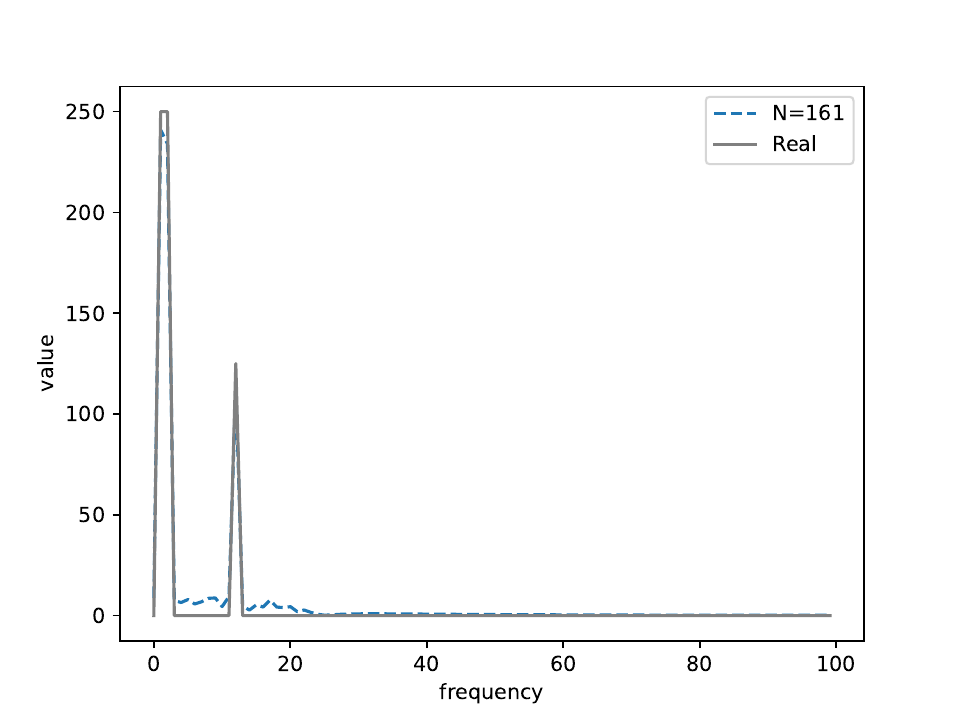}
\end{minipage}
}
\caption{A toy example of the effect of frequency parameter $N$. The data follows $f(t)=\sin{t}+\sin{2t}+0.5\sin{12t}$. (a) and (b) show the forecasting results on the time domain while (c) and (d) show the corresponding spectrum on frequency domain. The forecasts only show the low frequency information with $N=33$, but if we set $N=161$, which is a relatively large value, the forecasts are able to capture the high frequency information. }    
\label{fig:acc}    
\end{figure}

\subsection{Ineffectiveness of single Laplace decoder}
\begin{figure}[htb]
    \centering
    \includegraphics[width=\textwidth]{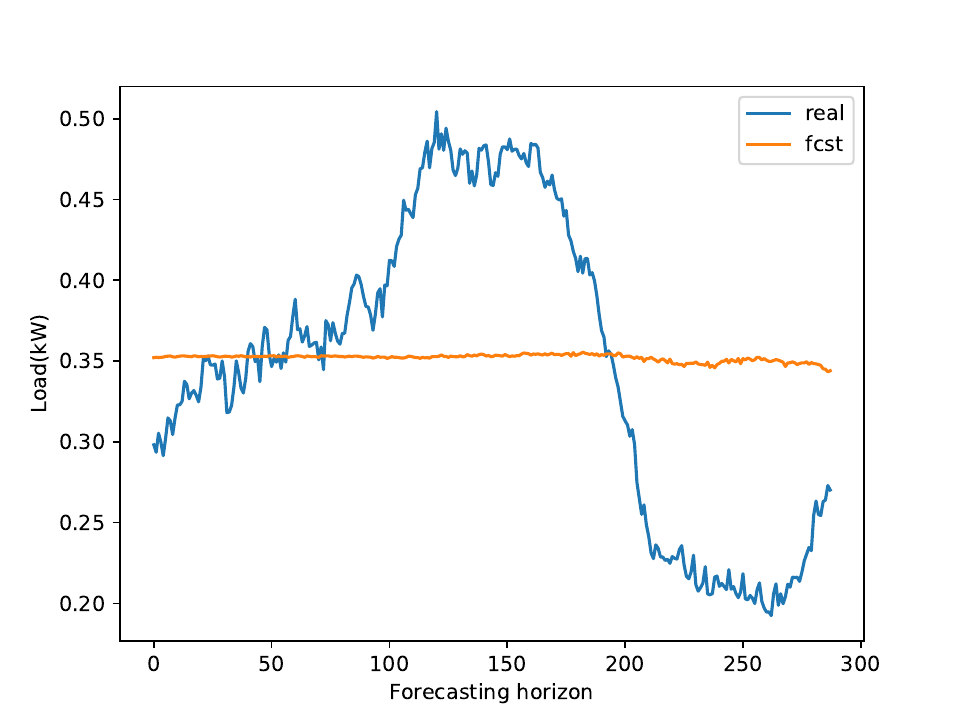}
    \caption{Ineffectiveness when only using one large Laplace decoder}
    \label{fig:supp_NL}
\end{figure}

We launched the experiment with one large Laplace decoder, i.e. Neural Laplace with a large enough frequency parameter $N$, trying to learn the whole spectrum of the energy data. However, the experiment turns out to be ineffective when only using one large Laplace decoder. The forecasting result can be observed in Fig.~\ref{fig:supp_NL}.

Therefore, when launching the benchmarks, we set the frequency parameter in the Neural Laplace as $N=33$ which is the same setting in the original code. In comparison, our proposed Hierarchical Laplace use multiple Laplace decoders to learn the frequency information in the energy data.

\subsection{Numerical experiment results}
We only show the visualization of the experiment results in the main text. This subsection will show the original numeric results of all the bar charts in the main text. It should be mentioned that the magnitudes of the load data and wind data differ a lot, therefore, the following results on the load dataset are scaled to demonstrate clearly.
Related tables include Table.~\ref{tab:load}-\ref{tab:day-ahead}.
\begin{table}[htb]
\centering
\caption{RMSE comparison on load dataset ($\times 10^{-2}$)}
\label{tab:load}
\begin{tabular}{@{}cccc@{}}
\toprule
Methods            & 5min                          & 15min                          & 60min                        \\ \midrule
LSTM        & 3.751 $\pm$ 0.142          & 3.809 $\pm$ 0.192          & 3.746 $\pm$ 0.159         \\
LSTM-BU     & 3.751 $\pm$ 0.142          & 3.708 $\pm$ 0.144          & 3.622 $\pm$ 0.146         \\
LSTM-OPT    & 3.597 $\pm$ 0.099          & 3.552 $\pm$ 0.100            & 3.463 $\pm$ 0.102         \\
MLP         & 3.591 $\pm$ 0.067          & 3.331 $\pm$ 0.114          & 3.130 $\pm$ 0.040           \\
MLP-BU      & 3.591 $\pm$ 0.067          & 3.547 $\pm$ 0.069          & 3.462 $\pm$ 0.071         \\
MLP-OPT     & 3.372 $\pm$ 0.060           & 3.325 $\pm$ 0.061          & 3.236 $\pm$ 0.062         \\
NL          & 3.638 $\pm$ 0.129          & 3.579 $\pm$ 0.164          & 3.503 $\pm$ 0.150          \\
NL-BU       & 3.638 $\pm$ 0.129          & 3.595 $\pm$ 0.131          & 3.501 $\pm$ 0.132         \\
NL-OPT      & 3.531 $\pm$ 0.099          & 3.486 $\pm$ 0.100            & 3.390 $\pm$ 0.101          \\
Persistence & 3.697 $\pm$ 0.000           & 3.615 $\pm$ 0.000            & 3.479 $\pm$ 0.000           \\
\textbf{HNL} & \textbf{3.428 $\pm$ 0.082} & \textbf{3.359 $\pm$ 0.079} & \textbf{3.360 $\pm$ 0.068} \\ \bottomrule
\end{tabular}
\end{table}

\begin{table}[htb]
\centering
\caption{RMSE comparison on wind power dataset}
\label{tab:wind}
\begin{tabular}{@{}cccc@{}}
\toprule
Methods     & 5min                       & 15min                      & 60min                      \\ \midrule
LSTM        & 5.189 $\pm$ 0.097          & 5.308 $\pm$ 0.196          & 5.403 $\pm$ 0.160           \\
LSTM-BU     & 5.189 $\pm$ 0.097          & 5.177 $\pm$ 0.097          & 5.118 $\pm$ 0.098          \\
LSTM-OPT    & 5.174 $\pm$ 0.086          & 5.162 $\pm$ 0.086          & 5.103 $\pm$ 0.087          \\
MLP         & 5.149 $\pm$ 0.091          & 4.928 $\pm$ 0.119          & 4.680 $\pm$ 0.087           \\
MLP-BU      & 5.149 $\pm$ 0.091          & 5.126 $\pm$ 0.090           & 5.063 $\pm$ 0.090           \\
MLP-OPT     & 5.028 $\pm$ 0.075          & 5.004 $\pm$ 0.073          & 4.941 $\pm$ 0.074          \\
NL          & 4.937 $\pm$ 0.380           & 4.972 $\pm$ 0.267          & 4.999 $\pm$ 0.349          \\
NL-BU       & 4.937 $\pm$ 0.380           & 4.930 $\pm$ 0.381           & 4.869 $\pm$ 0.384          \\
NL-OPT      & 4.813 $\pm$ 0.312          & 4.805 $\pm$ 0.312          & 4.743 $\pm$ 0.316          \\
Persistence & 6.596 $\pm$ 0.000            & 6.604 $\pm$ 0.000            & 6.640 $\pm$ 0.000             \\
\textbf{HNL} & \textbf{4.642 $\pm$ 0.252} & \textbf{4.617 $\pm$ 0.248} & \textbf{4.538 $\pm$ 0.251} \\ \bottomrule
\end{tabular}
\end{table}

\begin{table}[htb]
\centering
\caption{RMSE comparison on frequency domain}
\label{tab:freq_rmse}
\begin{tabular}{@{}ccc@{}}
\toprule
Method   & Load ($\times 10^{-2}$)      & Wind power         \\ \midrule
LSTM     & 58.771 $\pm$ 3.180 & 88.487 $\pm$ 2.71  \\
LSTM-OPT & 56.694 $\pm$ 2.297 & 88.548 $\pm$ 2.412 \\
MLP      & 55.202 $\pm$ 1.805 & 92.625 $\pm$ 2.526 \\
MLP-OPT  & 51.102 $\pm$ 1.508 & 90.352 $\pm$ 2.025 \\
NL       & 55.665 $\pm$ 2.403 & 82.677 $\pm$ 7.025 \\
NL-OPT   & 54.430 $\pm$ 1.602 & 80.987 $\pm$ 5.703 \\
HNL       & 50.003 $\pm$ 1.359 & 76.477 $\pm$ 4.576 \\ \bottomrule
\end{tabular}
\end{table}

\begin{table}[htb]
\centering
\caption{Consistency error comparison}
\label{tab:ce}
\begin{tabular}{@{}cccccc@{}}
\toprule
\multirow{2}{*}{Dataset}    & \multirow{2}{*}{Method} & \multicolumn{4}{c}{Consistency error}                             \\
                            &                         & 5min vs 15min  & 5min vs 60min  & 15min vs 60min & Total          \\ \midrule
\multirow{4}{*}{Load($\times 10^{-4}$)}       & LSTM                    & 5.973          & 7.882          & 6.598          & 20.452         \\
                            & MLP                     & 4.52           & 6.04           & 3.758          & 14.318         \\
                            & NL                      & 3.319          & 3.838          & 3.718          & 10.874         \\
                            & \textbf{HNL}             & \textbf{0.066} & \textbf{1.702} & \textbf{1.144} & \textbf{2.912} \\
\hdashline
\multirow{4}{*}{Wind power} & NL                      & 6.426          & 7.581          & 7.389          & 21.396         \\
                            & LSTM                    & 2.007          & 5.986          & 5.944          & 13.937         \\
                            & MLP                     & 2.084          & 3.932          & 2.499          & 8.515          \\
                            & \textbf{HNL}             & \textbf{0.036} & \textbf{0.294} & \textbf{0.177} & \textbf{0.507} \\ \bottomrule
\end{tabular}
\end{table}
\begin{table}[htb]
\centering
\caption{Operation cost comparison on day-ahead scheduling ($\times 10^5$ ¥)}
\label{tab:day-ahead}
\begin{tabular}{@{}cccc@{}}
\toprule
Method      & 5min                        & 15min                      & 60min                      \\ \midrule
LSTM        & 123.902 $\pm$ 10.309        & 41.63 $\pm$ 2.852          & 10.38 $\pm$ 0.797          \\
LSTM-BU     & 123.902 $\pm$ 10.309        & 42.635 $\pm$ 3.461         & 10.566 $\pm$ 0.865         \\
LSTM-OPT    & 117.941 $\pm$ 6.292         & 40.619 $\pm$ 2.127         & 10.056 $\pm$ 0.533         \\
MLP         & 123.426 $\pm$ 8.96          & 39.654 $\pm$ 3.745         & 9.127 $\pm$ 0.72           \\
MLP-BU      & 123.426 $\pm$ 8.96          & 42.464 $\pm$ 3.011         & 10.496 $\pm$ 0.757         \\
MLP-OPT     & 115.559 $\pm$ 6.105         & 39.813 $\pm$ 2.041         & 9.838 $\pm$ 0.517          \\
NL          & 119.641 $\pm$ 5.98          & 40.515 $\pm$ 2.91          & 9.977 $\pm$ 0.594          \\
NL-BU       & 119.641 $\pm$ 5.98          & 41.257 $\pm$ 2.017         & 10.204 $\pm$ 0.504         \\
NL-OPT      & 116.377 $\pm$ 4.509         & 40.166 $\pm$ 1.517         & 9.934 $\pm$ 0.371          \\
Persistence & 135.372 $\pm$ 0.0           & 46.49 $\pm$ 0.0            & 11.483 $\pm$ 0.0           \\
\textbf{HNL} & \textbf{107.526 $\pm$ 8.78} & \textbf{37.252 $\pm$ 2.99} & \textbf{9.785 $\pm$ 0.766} \\ \bottomrule
\end{tabular}
\end{table}

\subsection{Operation costs comparison under different penetration rates}
\begin{figure}[htb]
    \centering
    \includegraphics[width=0.71\textwidth]{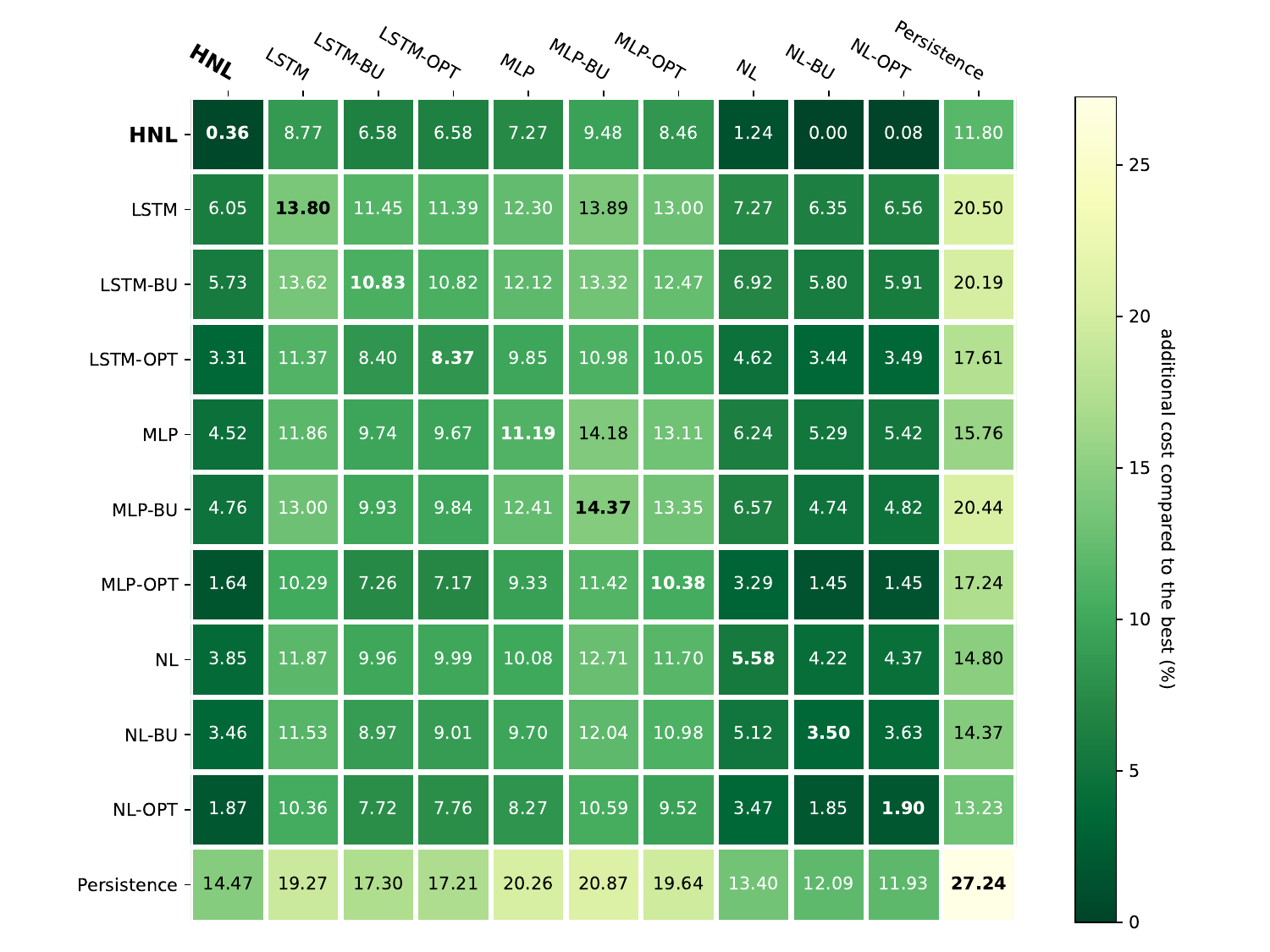}
    \caption{Comparison of operational costs for integrated scheduling (penetration rate: 20\%)}
    \label{fig:peneration_2}
\end{figure}

\begin{figure}[tb]
    \centering
    \includegraphics[width=0.71\textwidth]{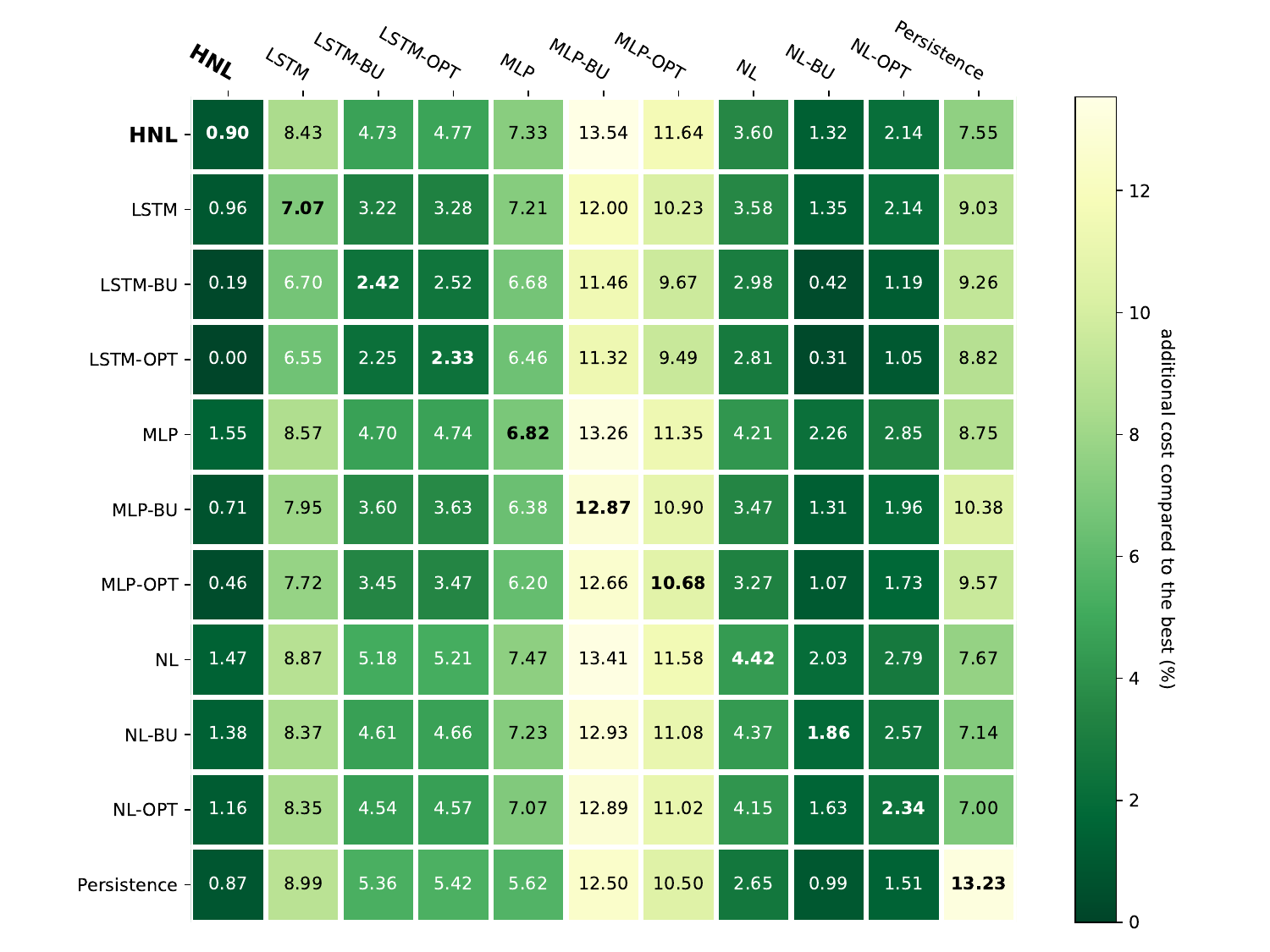}
    \caption{Comparison of operational costs for integrated scheduling (penetration rate: 80\%)}
    \label{fig:peneration_8}
\end{figure}

In the large-scale integrated scheduling problem, the penetration rate of wind power will determine which kind of forecasts (load or wind power) that plays a more critical role. Fig.~\ref{fig:peneration_2} and Fig.~\ref{fig:peneration_8} show the results under different penetration rate of wind power.

It can be seen that with the penetration rate get higher, the more distinction is shown among columns. This implies that the wind power forecasting methods gradually dominates the operation results when the penetration rate is higher. 

When the penetration rate is low (0.2), where load forecasting dominates, the lowest operation costs are achieved in the row (load forecasting methods) of HNL. Similarly, when the penetration rate is high (0.8), where wind power forecasting dominates, the lowest operation costs happens in the column (wind power forecasting methods) of HNL. It suggested that HNL shows dominant operational effects in all situations considered.

\end{appendices}

\end{document}